\documentclass[english,aps,twocolumn,prd,superscriptaddress,showpacs,preprintnumbers,amsmath,amssymb]{revtex4-1}

\usepackage{subfigure}
\usepackage{amsmath,amssymb}
\usepackage{graphicx}
\usepackage{rotating}
\usepackage{dsfont}
\usepackage{color}
\usepackage[dvips]{epsfig}     
\usepackage{multirow}
\definecolor{refcol}{RGB}{178,34,34}%{0,0,205}
\usepackage{microtype}
\usepackage[colorlinks,linkcolor=refcol,citecolor=refcol,urlcolor=refcol]{hyperref}
\usepackage{breakurl}
\usepackage{charter} 
\usepackage[charter]{mathdesign}

\usepackage[utf8]{inputenc}

\graphicspath{{./figures/}}

\def\di{\displaystyle}

\def\bg{\begin{eqnarray}\begin{array}{rcl}\displaystyle}
\def\eg{\end{array} &\di    &\di   \end{eqnarray}}
\def\bm#1{\begin{eqnarray}\begin{array}{#1}\di}
\def\bmo#1{\begin{eqnarray*}\begin{array}{#1}\di}
\def\bml#1#2{\begin{eqnarray}\begin{array}{#1}\label{#2}\di}
\def\bgo{\begin{eqnarray*}\begin{array}{rcl}\displaystyle}
\def\ego{\end{array} &\di    &\di \nonumber  \end{eqnarray*}}

\def\btensor#1#2{\renew\left#1\begin{array}{#2}\di}
\def\brtensor#1#2#3{\ren#3\left#1\begin{array}{#2}}
\def\botensor#1#2{\renew\left#1\begin{array}{#2}}
\def\etensor#1{\end{array}\right#1}

\def\eq#1{(\ref{#1})}
\def\Eq#1{Eq.~(\ref{#1})}

\def\det{{\rm det}}

\def\s0#1#2{\mbox{\small{$ \frac{#1}{#2} $}}}
\def\0#1#2{\frac{#1}{#2}}

\def\ren#1{\renewcommand{\arraystretch}{#1}}

\def\renew{\renewcommand{\arraystretch}{1}}

\definecolor{blue}{rgb}{0,0,1}

\definecolor{green}{rgb}{0,1,0}

\definecolor{red}{rgb}{1,0,0}

\definecolor{bjcol}{rgb}{1,.44,0.13}

\newcommand{\Tr}{\mathrm{Tr}}

\newcommand{\tr}{\mathrm{tr}}

\newcommand{\be}{\begin{eqnarray}}
\newcommand{\ee}{\end{eqnarray}}

\usepackage{babel}
\makeatother

\begin{document}

\title{Fluctuation-induced modifications of the phase structure in
  $(2\!+\!1)$-flavor QCD}
\author{Fabian Rennecke}
\email[E-mail: ]{fabian.rennecke@theo.physik.uni-giessen.de}
\affiliation{Institut f\"ur Theoretische
  Physik, Justus-Liebig-Universit\"at Gie\ss en, Heinrich-Buff-Ring 16, 35392 Gie\ss
  en,Germany}
\affiliation{Institut
 f\"{u}r Theoretische Physik, Universit\"{a}t Heidelberg,
 Philosophenweg 16, 69120 Heidelberg, Germany}

\author{Bernd-Jochen Schaefer}
\email[E-mail: ]{bernd-jochen.schaefer@theo.physik.uni-giessen.de}
\affiliation{Institut f\"ur Theoretische
  Physik, Justus-Liebig-Universit\"at Gie\ss en, Heinrich-Buff-Ring 16, 35392 Gie\ss
  en,Germany}

\begin{abstract}
  The low-energy sector of QCD with $N_f = 2\!+\!1$ dynamical quark
  flavors at non-vanishing chemical potential and temperature is
  studied with a non-perturbative functional renormalization group
  method. The analysis is performed in different truncations in order
  to explore fluctuation-induced modifications of the quark-meson
  correlations as well as quark and meson propagators on the chiral
  phase transition of QCD. Depending on the chosen truncation
  significant quantitative implications on the phase transition are
  found. In the chirally symmetric phase, the quark flavor composition
  of the pseudoscalar $(\eta,\eta^{\prime})$-meson complex turns out
  to be drastically sensitive to fluctuation-induced modifications in
  the presence of the axial $U(1)_A$ anomaly. This has important phenomenological consequences for the assignment of chiral partners to these mesons.
\end{abstract}

\maketitle

%%%%%%%%%%%%%%%%%%%%%%%%%%%%%%%%%%%%%%%%%%%%%%%%%%%%%%%%%%%%%%
%%%%%%%%%%%%%%%%%%%%%%%%%%%%%%%%%%%%%%%%%%%%%%%%%%%%%%%%%%%%%%
\section{Introduction}\label{sec:intro}

The phase structure of QCD remains to be one of the most important yet
unresolved problems of heavy-ion physics. Owing in particular to
lattice gauge theory simulations, the region of very small density
seems to be well understood. A crossover transition from a confined
phase with explicitly and spontaneously broken chiral symmetry around
the pseudocritical temperature $T_c \!\approx\! 155\,\text{MeV}$ to
the quark-gluon plasma phase takes place at vanishing quark chemical
potential \cite{Aoki:2006we, Borsanyi:2010bp,
  Petreczky:2012rq}. However, the notorious sign problem renders
lattice simulations unreliable at larger densities. While much
progress in circumventing the sign problem has been made in recent
years, see e.g. \cite{Aarts:2015tyj}, a full resolution in the near
future is highly unlikely. However, present and future heavy-ion
collider experiments such as the Beam Energy Scan at RHIC
\cite{Adamczyk:2013dal, Adamczyk:2014fia}, the NA61/SHINE experiment
at the SPS at CERN \cite{Gazdzicki:995681} or the CBM experiment at
FAIR \cite{Friman:2011zz} as well as NICA at JINR \cite{Kekelidze2012}
start to investigate also the large density region of QCD. It is
therefore indispensable to develop reliable descriptions of QCD at
finite density.

Functional continuum methods such as the functional renormalization
group (FRG) \cite{Braun:2009gm, Mitter:2014wpa, Cyrol:2016tym} or
Dyson-Schwinger equations (DSE) \cite{Alkofer:2000wg, Fischer:2006ub,
  Fischer:2014ata} provide powerful non-perturbative tools to describe
QCD from first principles also at finite density. In both cases, a
solution of QCD corresponds to the solution of an infinite tower of
coupled partial differential equations. In order to reduce the system
to a manageable size, truncation schemes have to be
devised. Systematic extensions of a given truncation allow for
systematic error estimates and can be utilized to identify
quantitatively precise approximation schemes.

One of the biggest challenges is to capture the drastic change of
relevant degrees of freedom at the phase boundary. While quarks and
gluons are the relevant objects above $T_c$, hadrons determine the
system for low temperatures.  Owing to asymptotic freedom the
underlying idea is that a rising gauge coupling towards lower scales
generates effective four-quark (and higher) interactions.  At the
scale of chiral symmetry breaking, these interactions develop poles in
the corresponding quark-antiquark scattering channels which signals
the formation of mesonic bound states. At lower energy scales they
then become the dominant degrees of freedom.  Within the FRG a
practical description of this dynamical transition from quarks and
gluons to hadrons is provided by dynamical hadronization, see
\cite{Gies:2001nw, Gies:2002hq, Pawlowski:2005xe, Floerchinger:2009uf,
  Braun:2014ata} and \cite{Braun:2008pi, Mitter:2014wpa, Braun:2014ata,
  Rennecke:2015eba} for explicit QCD applications. Here we restrict
ourselves to the low-energy sector and assume that the gauge sector is
fully integrated out at these scales.
How such quark-meson-type models emerge dynamically in the low-energy
sector of QCD under these assumptions is well understood and explained
e.g.~in \cite{Schaefer:2006sr, Braun:2009gm,
  Kondo:2010ts, Herbst:2010rf, Herbst:2013ail, Haas:2013qwp,
  Herbst:2013ufa, Mitter:2014wpa, Braun:2014ata}. However, the precise scale of the gluon
decoupling is still under debate which we will pick up in
Sec.~\ref{sec:ini}. For the present purpose, it is sufficient to
assume the decoupling of the gluons below a certain UV scale
$\Lambda$.

In the present work, we concentrate on the hadronic low-energy sector
of QCD with $N_f =2+1$ dynamical quark flavors at finite temperature
and chemical potential. Our main question is how quark and meson
fluctuations influence the chiral phase boundary,
cf.~e.g.~\cite{Schaefer:2011pn}. This allows us to gain valuable
insights into the impact of these fluctuation on the critical endpoint
(CEP) as well as the physics in the vicinity of the chiral phase
boundary. The axial $U(1)_A$ anomaly as well as mesonic
self-interactions in terms of the full effective meson potential have
been investigated with the FRG in \cite{Mitter:2013fxa}. It was shown
that fluctuation-induced effects beyond mean-field significantly
weaken the chiral phase transition compared to standard mean-field
studies such that the CEP is moved to small temperature and large
chemical potential. In addition, the location of the CEP also strongly
depends on the presence of the axial anomaly when meson fluctuations
are taken into account. In \cite{Mitter:2013fxa} quarks and mesons
have been assumed to obey their classical dispersion relations, i.e.,
the propagators have the classical momentum dependence and quantum
corrections to quark-meson correlations have been neglected. However,
fluctuation-induced corrections to both the dispersion relations and
quark-meson correlations can have a significant impact on the chiral
phase boundary as found recently for low-energy QCD with two flavors
\cite{Pawlowski:2014zaa} and play a crucial role in the transition
from the quark-gluon to the hadronic regime in two-flavor QCD
\cite{Braun:2014ata, Rennecke:2015eba}.  This motivates us to include
these corrections of the low-energy sector of $(2+1)$-flavor QCD in
the present study by allowing for running quark and meson wave
function renormalizations as well as running quark-meson Yukawa
couplings. In this way we substantially extend our previous works in
\cite{Mitter:2013fxa, Pawlowski:2014zaa}.

Phenomenologically, chiral symmetry breaking manifests itself in the
absence of parity-doublets in the experimentally observed spectrum of
low-mass mesons. On the other hand, chiral symmetry restoration is
signaled by the degeneration of the masses of chiral partners. In
addition to the total and orbital angular momentum, also the quark
composition determines which mesons eventually form chiral partners if
chiral symmetry is restored. In particular the quark composition of
mesons which cannot be described as simple flavor-eigenstates, can
potentially be intricate. A relevant example is the pseudoscalar
$\eta$- and $\eta^\prime$-meson complex, whose quark composition is
described by the pseudoscalar mixing angle $\varphi_p$. This angle
depends on the microscopic details of the theory as well as on
temperature and density. This dependence has been studied within
  Nambu--Jona-Lasinio, linear sigma and quark-meson models mostly in
  mean-field approximation \cite{Lenaghan:2000ey, Costa:2005cz,
    Schaefer:2008hk, Tiwari:2013pg, Kovacs:2016juc}. Here, we go
beyond such approximations and study the impact of quark and meson
fluctuations on the mixing angles at finite temperature and density.

The paper is organized as follows: We develop a low-energy model for
the hadronic sector of 2+1-flavor QCD, which includes the above
discussed effects in Sec.~\ref{sec:model}. In the following Section
\ref{sec:flucts} the incorporation of non-perturbative quantum and
thermal fluctuation by means of the functional renormalization group
is depicted. The corresponding flow equations for the effective
potential, wave function renormalization as well as Yukawa couplings
are derived in detail. Our results are
presented in Sec.~\ref{sec:res}. After discussing the initial
conditions for the flow equations, we first compare the chiral phase
structure including a critical endpoint in different truncations. A
brief discussion on related systematic errors will be given in
Sec.~\ref{sec:pb}.  The quark composition of scalar and pseudoscalar
mesons and its relation to the axial anomaly is addressed. We
summarize our results and give an outlook in Sec.~\ref{sec:
  concl}. Computational details are provided in the four appendices.

%%%%%%%%%%%%%%%%%%%%%%%%%%%%%%%%%%%%%%%%%%%%%%%%%%%%%%%%%%%%%%
%%%%%%%%%%%%%%%%%%%%%%%%%%%%%%%%%%%%%%%%%%%%%%%%%%%%%%%%%%%%%%
\section{Effective Model for $N_f =2+1$ QCD at Low Energies}\label{sec:model}

As argued in the introduction the low-energy sector of $N_f=2+1$ quark
flavor QCD can be described by a linear sigma model with dynamical (constituent) quarks also known as a quark-meson model.
As explained in more detail in Sec.~\ref{sec:frg}, we include quantum
fluctuations by means of the FRG. Since quantum effects in this
approach are driven by off-shell fluctuations of the fields, the
lightest mesons give rise to the dominant quantum corrections. 
The dynamics of low-energy QCD are captured decisively by the
inclusion of the pseudoscalar meson nonets and chiral symmetry
requires the inclusion of the scalar meson nonet.  Hence, we consider
the lightest scalar and pseudoscalar meson nonets and the three
lightest (constituent) quarks $u$, $d$ and $s$. A simplified variant of the resulting effective
model setup has been studied in \cite{Mitter:2013fxa}. Here, we go
beyond this work by allowing for fluctuation-induced corrections to
the classical dispersion relations of quarks and mesons as well as to
quark-meson scattering for the first time.

This leads us to the RG scale-dependent effective action
\begin{align}\label{eq:trunc}
\begin{split}
\Gamma_k &= \int_x \biggl\{i \bar  q Z_{q,k}  \bigl( \gamma_\mu\partial_\mu + \gamma_0\mu\bigr)\, q + i \bar q\, h_{q,k} \!\cdot\!  \Sigma_{5}  q\\
&\quad  + \tr\,\bigl(\!Z_{\Sigma,k} \partial_\mu\Sigma\!\cdot\! \partial_\mu\Sigma^\dagger\bigr) +\tilde U_k(\Sigma,\Sigma^\dagger)
\biggr\}\,,
\end{split}
\end{align}
with the flavor independent quark chemical potential $\mu$ and the
finite temperature $T$ introduced via the Matsubara formalism such
that the integral reads $\int_x = \int_0^{1/T}\!dx_0\int\!d^3x$. We
assume isospin symmetry in the light quark flavor sector and label the
quark fields by $q=(l,l,s)$ with the light quarks $l \equiv u=d$ and
the strange quark $s$.  As denoted, all parameters depend on the RG scale $k$. The
quark wave function renormalization $Z_{q,k}$ encodes the
fluctuation-induced modifications of the dispersion relation of the
quark. The running Yukawa coupling $h_{q,k}$ couples mesons to
quark-antiquark pairs. Both $Z_{q,k}$ and $h_{q,k}$ are matrices in
flavor space and read for the isospin symmetric theory,
\begin{align}\label{eq:zandh}
Z_{q,k} = \begin{pmatrix} Z_{l,k} & 0 & 0 \\ 0 & Z_{l,k} & 0 \\ 0 & 0 & Z_{s,k} \end{pmatrix}\,, \quad
h_{q,k} = \begin{pmatrix} h_{l,k} & h_{l,k} & h_{ls,k} \\ h_{l,k} & h_{l,k} & h_{ls,k} \\ h_{sl,k} & h_{sl,k} & h_{s,k} \end{pmatrix}\,.
\end{align}
The meson field $\Sigma$ includes the scalar and pseudoscalar mesons
$\sigma_a$ and $\pi_a$ in the adjoint representation of $U(N_f)$,
\begin{align}\label{eq:Sigma}
\Sigma = T_a (\sigma_a + i \pi_a)\,,\quad a=0,\ldots,8
\end{align}
where $T_1,\dots,T_{N_f^2-1}$ are the generators of $SU(N_f)$ and
$T_0 = \frac{1}{\sqrt{2N_f}} \mathds{1}_{N_f\times N_f}$. For $N_f=3$
the generators are given by the usual Gell-Mann matrices
$\lambda_a = 2 T_a$. Additionally, the $\Sigma_5$ field couples to the quarks in a chirally
symmetric manner and reads
\begin{align}
\Sigma_5 = T_a (\sigma_a + i \gamma_5 \pi_a)\,.
\end{align}
The meson effective potential $\tilde U_k(\Sigma,\Sigma^\dagger)$ in \eqref{eq:trunc}
encodes all momentum-independent mesonic correlation functions. It can be decomposed into a
chirally symmetric part and various contributions that break subgroups
of the underlying $U(3) \!\times\! U(3)$ chiral symmetry. The chirally
invariant part of this potential has to be a function of the
$U(3)\!\times\! U(3)$ invariants $\rho_1$ and $\tilde\rho_2$, which
are defined as
\begin{align}
\begin{split}
\rho_1 &= \tr\, \left( \Sigma\!\cdot\!\Sigma^\dagger  \right)\,,\\
\tilde\rho_2 &= \tr\,\Bigl(\!\Sigma\!\cdot\!\Sigma^\dagger- \frac{1}{2} \rho_1 \mathds{1}_{3\times3}\Bigr)^2\,.
\end{split}
\end{align}
In general, $\tilde U_k$ could additionally depend on the third chiral
invariant
$\tilde\rho_3 = \tr\,(\Sigma\cdot\Sigma^\dagger- \frac{1}{2} \rho_1
\mathds{1}_{3\times3})^3$,
which we drop here for the sake of simplicity. In order to capture the
anomalous breaking of the axial $U(1)_A$ symmetry, i.e. the axial
anomaly, we furthermore introduce the instanton induced
Kobayashi-Maskawa-'t Hooft (KMT) determinant \cite{Kobayashi:1970ji,
  'tHooft:1976up, 'tHooft:1976fv},
\begin{align}
\xi = \det\,(\Sigma+\Sigma^\dagger)\,,
\end{align}
to the potential, which constitues the lowest-order of a
$U(1)_A$-symmetry breaking operator.  The non-vanishing masses of the
Goldstone bosons, which are directly linked to the finite current
quark masses, are incorporated by an explicit chiral symmetry breaking
via
\begin{align}\label{eq:esb}
\tr\,\bigl[\!T_a j_a (\Sigma+\Sigma^\dagger)\bigr] = j_a \sigma_a\,.
\end{align}
Due to the spontaneous chiral symmetry breaking in the vacuum, a finite
vacuum expectation value (VEV) of the $\Sigma$
field is generated which should carry the quantum numbers of the
vacuum. This results in only non-vanishing explicit symmetry
breaking parameters that correspond to the diagonal $U(3)$-generators
$T_0, T_3$ and $T_8$. Otherwise, unphysical charged chiral condensates
would be generated. 
For the assumed light isospin symmetry this yields
$j_0, j_8 \neq 0$ and $j_3 =0$. Consequently, only the scalar fields $\sigma_0$
and $\sigma_8$ assume non-vanishing VEVs which are determined by
minimization of the effective potential.

In summary, the effective potential has the following structure 
\begin{align}\label{eq:effpot}
\tilde U_k(\Sigma,\Sigma^\dagger) = U_k(\rho_1,\tilde \rho_2)- c_k\xi - j_L\sigma_L - j_S \sigma_S\,,
\end{align}
wherein the chirally symmetric part of the potential,
$U_k(\rho_1,\tilde \rho_2)$, is still an arbitrary function of the
chiral invariants $\rho_1$ and $\tilde\rho_2$. In practice, we will
compute it numerically by using the two-dimensional generalization of
the fixed background Taylor expansion that was put forward in
\cite{Pawlowski:2014zaa}. A detailed discussion of the numerical
implementation can be found in App.~\ref{app:taylor}. Note that we
have expressed the explicit symmetry breaking terms in
\Eq{eq:effpot} in the light-strange (LS) basis as explained below.

The explicit symmetry breaking terms $j_{L/S}$ are directly related to
the quark masses $m_{q,k}$ via
\begin{align}
j_L = \frac{2 \nu_{L,k}}{h_{l,k}} m_{l,k}\,,\qquad j_S = \frac{\sqrt{2} \nu_{S,k}}{h_{s,k}} m_{s,k}\,,
\end{align}
with the coefficients
$\nu_{L/S,k} = \sigma_{L/S}^{-1}\partial_{\sigma_{L/S}}(U_k-c_k\xi)$,
where the expression inside the bracket is evaluated at the minimum of
$\tilde{U}_k$. Without loss of generality, we choose $j_{L/S}$ to be
$k$-independent.  The quark masses themselves are related to the
mesonic VEVs and the Yukawa couplings by
\begin{align}\label{eq:qmasses}
m_{l,k} = \frac{h_{l,k}}{2} \sigma_L\,,\qquad m_{s,k} = \frac{h_{s,k}}{\sqrt{2}} \sigma_S\,.
\end{align}
The (bare) pion and kaon decay constants $f_\pi$ and $f_K$ are related to the mesonic VEVs by \cite{Lenaghan:2000ey}
\begin{align}
f_\pi = \sigma_L\,,\qquad f_K = \frac{\sigma_L+\sqrt{2}\sigma_S}{2}\,.
\end{align}
Later, the mesonic VEVs become also $k$-dependent which we suppress
here. In terms of the physical scalar and pseudoscalar meson fields,
$\Sigma$ reads
\begin{widetext}
\begin{align}\label{eq:sigmama}
\Sigma = \frac{1}{\sqrt{2}}
\begin{pmatrix}
\frac{1}{\sqrt{2}}\left(\sigma_L+a_0^0+i\eta_L+i\pi^0\right) & a_0^- + i \pi^- & \kappa^-+i K^- \\
a_0^+ + i \pi^+ &  \frac{1}{\sqrt{2}}\left(\sigma_L - a_0^0+i\eta_L - i\pi^0\right) & \kappa^0+i K^0 \\
 \kappa^++i K^+ & \bar\kappa^0+i \bar K^0 & \frac{1}{\sqrt{2}}\left(\sigma_S+i\eta_S\right)
\end{pmatrix}\,,
\end{align}
\end{widetext}
where we have employed the light-nonstrange--strange (LS) basis, in
which the original scalar $\sigma$ (pseudoscalar $\eta$) fields
decompose into mesons made of only light-nonstrange quarks
$\sigma_L,\,\eta_L$ and strange quarks $\sigma_S,\,\eta_S$
respectively. Hence, they are flavor eigenstates. The physical states
are the mass eigenstates which are obtained from diagonalizing the
Hessian
$H_{ij} = \partial_{\phi_i}\partial_{\phi_j} \tilde
U(\Sigma,\Sigma^\dagger)$,
where the fields $\phi_i$ comprises all physical mesons in the LS
basis as in \Eq{eq:sigmama}.
Note that there are no two-point functions that mix scalar and
pseudoscalar fields. Thus, $H_{ij}$ assumes a block-diagonal form with
one scalar and one pseudoscalar block. The off-diagonal elements
in each block are the $\sigma_L,\sigma_S$ and $\eta_L,\eta_S$
entries. Since the Hessian is furthermore symmetric, we can
diagonalize it by introducing the scalar and pseudoscalar mixing
angles $\varphi_s$ and $\varphi_p$ which rotate the LS-fields into the
known physical states
\begin{align}\label{eq:plstrafo}
\begin{split}
\begin{pmatrix}f_0 \\ \sigma \end{pmatrix} &= \begin{pmatrix} \cos\varphi_s & - \sin\varphi_s \\ \sin\varphi_s & \cos\varphi_s \end{pmatrix} \begin{pmatrix}\sigma_L \\ \sigma_S \end{pmatrix}\,,\\
\begin{pmatrix}\eta \\ \eta^\prime \end{pmatrix} &= \begin{pmatrix} \cos\varphi_p & - \sin\varphi_p \\ \sin\varphi_p & \cos\varphi_p \end{pmatrix} \begin{pmatrix}\eta_L \\ \eta_S \end{pmatrix}\,.
\end{split}
\end{align}
All other fields in \Eq{eq:sigmama} are already mass
eigenstates.

Furthermore, the fields in the 
singlet-octet (08) basis as used in \Eq{eq:Sigma} can be transformed
into the LS basis by
\begin{align}\label{eq:ls08}
\begin{pmatrix}\phi_L \\ \phi_S \end{pmatrix} = \frac{1}{\sqrt{3}}\begin{pmatrix} 1 & \sqrt{2} \\ -\sqrt{2} & 1 \end{pmatrix} \begin{pmatrix}\phi_8 \\ \phi_0 \end{pmatrix}\,.
\end{align}
For more details on the mixing angles we defer to
App.~\ref{app:mixing}.  In summary, the vector of all physical mesons
then is
\begin{align}\label{eq:mesons}
\begin{split}
\phi =(&f_0, a_0^0, a_0^+, a_0^-,\kappa^+,\kappa^-,\kappa^0,\bar\kappa^0,\sigma,\\
&\eta, \pi^0, \pi^+, \pi^-,K^+,K^-,K^0,\bar K^0,\eta^\prime)\,.
\end{split}
\end{align}
In the present isospin symmetric setup only the $f_0$ and $\sigma$
assume a non-vanishing vacuum expectation value. These are directly
linked to the light and strange quark condensates which are, in turn,
related to the vacuum expectation values of $\sigma_L$ and
$\sigma_S$. We therefore conveniently define the VEV of $\phi$ as
\begin{align}
\phi_{0} \equiv (\sigma_{L},\sigma_{S}, 0, \ldots, 0)\,.
\end{align}

%%%%%%%%%%%%%%%%%%%%%%%%%%%%%%%%%%%%%%%%%%%%%%%%%%%%%%%%%%%%%%
%%%%%%%%%%%%%%%%%%%%%%%%%%%%%%%%%%%%%%%%%%%%%%%%%%%%%%%%%%%%%%
\section{Fluctuations}\label{sec:flucts}

As already mentioned the present work aims at fluctuation-induced
effects in the low-energy sector of QCD which are particularly
relevant in the vicinity of the chiral phase transition where
long-range correlations occur.  In the FRG framework quantum
fluctuations result in a running of all parameters of the effective
average action $\Gamma_k$ in \Eq{eq:trunc}. In the following, we will
give a brief outline of the used FRG to incorporate fluctuations
here. This is followed by a discussion on the corresponding beta
functions, or flow equations of the parameters of our truncation. For reviews of the FRG we refer the reader to \cite{Berges:2000ew, Pawlowski:2005xe, Gies:2006wv, Schaefer:2006sr, Delamotte:2007pf, Rosten:2010vm, Braun:2011pp}.
\vspace{-0.2cm}

%%%%%%%%%%%%%%%%%%%%
\subsection{The Functional Renormalization Group}\label{sec:frg}
\vspace{-0.1cm}

The underlying idea of K. G. Wilson's version of the renormalization
group is to integrate out quantum fluctuations successively . In the
present case, one begins with the microscopic action
$\Gamma_{k=\Lambda}$ at some large initial momentum scale $\Lambda$ in
the ultraviolet (UV). By lowering the RG scale $k$, quantum
fluctuations are successively integrated out until one arrives at the
full macroscopic quantum effective action $\Gamma \equiv \Gamma_{k=0}$ at
$k=0$. Ideally, one starts in the perturbative regime where the
initial effective action $\Gamma_{k=\Lambda}$ is given by the
well-known microscopic action of QCD. In our truncation we assume that
gluon degrees of freedom are already integrated out which fixes the
initial scale by the scale of the gluon decoupling and is therefore
directly linked to the gluon mass gap.
Based on results in Landau gauge QCD \cite{Fischer:2008uz, Maas:2011se}, this yields UV
scale choices of the order $\Lambda \lesssim 1\,\text{GeV}$.

A convenient realization of Wilson's RG idea is given in terms of a
functional differential equation for the 1PI effective average action,
the Wetterich equation \cite{Wetterich:1992yh}.
For the present $2+1$-flavor quark-meson model truncation with two
meson nonets and two light and one strange quark flavor the evolution
equation of the scale dependent effective average action $\Gamma_k$
reads explicitly
\begin{align}\label{eq:fleq}
\begin{split}
\partial_t \Gamma_k &= \frac{1}{2}\sum_{i=1}^{2N_f^2} \Tr \left(\!G_{\phi_i,k} \!\cdot \partial_t R_k^{\phi_i}\right)\\
&\quad - 2\Tr \left(G_{l,k} \!\cdot \partial_t R_k^l\right)- \Tr \left( G_{s,k} \!\cdot \partial_t R_k^s\right)\,,
\end{split}
\end{align}
where $\partial_t = k\frac{ d}{ d k}$ denotes the logarithmic scale
derivative. The trace runs over all discrete and continuous indices,
i.e. color, spinor and the loop momenta and/or frequencies
respectively. The sum in the first line is over all $2 N_f^2$ scalar
and pseudoscalar mesons, cf.~\Eq{eq:mesons}. $G_{\phi_{i},k}$ denotes the
generalized meson and quark propagators
\begin{align}
\begin{split}
G_{\phi_i,k}[\Phi] &= \left(\frac{\delta^2\Gamma_k[\Phi]}{\delta\phi_i(-p)\delta\phi_i(p)} + R_k^{\phi_i} \right)^{-1}\,,\\
G_{q,k}[\Phi] &= \left(\frac{\delta^2\Gamma_k[\Phi]}{\delta q(-p)\delta\bar q(p)} + R_k^{q} \right)^{-1}\,,
\end{split}
\end{align}
with the super field $\Phi=(\phi, q, \bar q)$. 
Evaluated on the equations of motion at $k=0$, i.e., on the vacuum
expectation value $\Phi_0 = (\phi_0,0,0)$, these generalized
propagators reduce to the full propagators. The scale-dependent IR
regulators $R_k^{\phi_i}$ can be understood as momentum-dependent
squared masses that suppress the infrared modes of the field
$\phi_i$. In addition, the terms $\partial_t R_k^{\phi_i}$ in
\Eq{eq:fleq} also ensure UV-regularity. Their definitions (see
\Eq{eq:regs}) and further details are collected in
App.~\ref{app:rucou}.

Inserting the truncation for the effective average action $\Gamma_k$
in \Eq{eq:trunc} in the flow equation \eq{eq:fleq} yields a closed set
of coupled differential equations for all scale-dependent parameters
of $\Gamma_k$.  Note that the physical parameters are RG-scale
dependent but invariant under general reparameterizations. 
In the present case, this is achieved by an appropriate rescaling with
the wave function renormalizations. The physical quark and meson
masses are then given by
\begin{align}\label{eq:renma}
M_{q,k} = \frac{m_{q,k}}{Z_{q,k}}\,,\qquad M_{\phi_i,k} = \frac{m_{\phi_{i},k}}{Z_{\phi_i,k}^{1/2}}\,,
\end{align}
where the meson masses $m_{\phi_i,k}$ are the eigenvalues of the
Hessian of the effective potential. The explicit expressions are given
in App.~\ref{app:mesmc}.

For simplicity, however, we assume that all mesons have the same
anomalous dimension $Z_{\phi,k}\equiv Z_{\phi_i,k}$, i.e., we choose
$Z_{\Sigma,k} = \text{diag}\,(Z_{\phi,k},\,Z_{\phi,k},\,Z_{\phi,k})$
in \Eq{eq:trunc}. Accordingly, the scale-dependent physical VEV of the
fields is given by
\begin{align}
\label{eq:1}
\bar \phi_0 = (\bar\sigma_{L},\bar\sigma_{S}, 0, \ldots, 0) = Z_{\phi,k}^{1/2} \phi_0\,,
\end{align}
where the index $k$ on the VEV's has been
suppressed. In the following we will use the bar notation over a
coupling to denote its rescaling with the wave function
renormalization. The running Yukawa coupling is
\begin{align}\label{eq:ruh}
\bar h_{q,k} = \frac{h_{q,k}}{Z_{q,k} Z_{\phi,k}^{1/2}}\,.
\end{align}
The chiral invariants and the KMT determinant are then
\begin{align}
\bar\rho_1 = Z_{\phi,k}\rho_1\,,\quad \bar{\tilde\rho}_2 = Z_{\phi,k}^2\tilde{\rho}_2\,,\quad \bar\xi = Z_{\phi,k}^{3/2}\xi\,,
\end{align}
and the effective potential is fixed from the requirement of reparametrization invariance,
\begin{align}
\bar U_k(\bar\rho_1,\bar{\tilde\rho}_2) = U_k(\rho_1,\tilde\rho_2)\,.
\end{align}
The consequence of reparametrization invariance in the present case is
that the wave function renormalizations do not enter the flow
equations directly, but only through the corresponding anomalous
dimensions
\begin{align}\label{eq:etadef}
\eta_{\Phi,k} = -\frac{\partial_t Z_{\Phi,k}}{Z_{\Phi,k}}\,.
\end{align}
Thus, introducing wave function renormalizations in this way implies
that it is not necessary to solve their flow equation explicitly. One
only needs to know the anomalous dimensions for the solution of the
present truncation. The anomalous dimensions, in turn, can be computed
analytically.

In the following, we briefly discuss the structure of the
flow equations for the effective potential, Yukawa couplings and
anomalous dimension.

%%%%%%%%%%%%%%%%%%%%
\subsection{Flow of the Effective Potential}\label{sec:pot}

All physical information is extracted from the effective potential
evaluated at the minimum at $k=0$. The meson fields assume their
vacuum expectation value at this point. As discussed above, the VEV
$\phi_0$ has two non-vanishing components which are most conveniently
expressed in the LS-basis, $\sigma_{L,k=0}$ and $\sigma_{S,k=0}$. With
a slight abuse of terminology, we refer to them as the light and
strange condensates \footnote{Strictly speaking, there is only a
  one-to-one correspondence between the chiral condensates and the
  minimum of the effective potential.}. They are obtained from the
simultaneous solution of
\begin{align}
\begin{split}
\partial_{\sigma_L} \tilde U_k(\Sigma,\Sigma^\dagger)\big|_{\phi_0} = 0\,,\quad \text{and}\quad \partial_{\sigma_S} \tilde U_k(\Sigma,\Sigma^\dagger)\big|_{\phi_0} = 0\,.
\end{split}
\end{align}
The flow of the effective potential is obtained from the
scale-dependent effective action for constant fields. We find for the
flow of the chirally invariant part of the potential in terms of the
physical fields:
\begin{align}\label{eq:uflow}
 &\partial_{t} U_k(\rho_1,\tilde\rho_2) = \\ \nonumber
&\quad \frac{k^4}{4\pi^2} \bigg\{ l_0^{(B)}(\bar m_{f_0,k}^2,\eta_{\phi,k}) + 3 l_0^{(B)}(\bar m_{a_0,k}^2,\eta_{\phi,k})\\ \nonumber
&\quad +4l_0^{(B)}(\bar m_{\kappa,k}^2,\eta_{\phi,k})+ l_0^{(B)}(\bar m_{\sigma,k}^2,\eta_{\phi,k}) + l_0^{(B)}(\bar m_{\eta,k}^2,\eta_{\phi,k})\\ \nonumber
&\quad +3 l_0^{(B)}(\bar m_{\pi,k}^2,\eta_{\phi,k}) +4 l_0^{(B)}(\bar m_{K,k}^2,\eta_{\phi,k}) + l_0^{(B)}(\bar m_{\eta^\prime,k}^2,\eta_{\phi,k}) \\ \nonumber
&\quad \qquad- 4N_c  \Big[ 2 l_0^{(F)}(\bar m_{l,k},\eta_{l,k}) +
  l_0^{(F)}(\bar m_{s,k},\eta_{s,k}) \Big] \bigg\}\,.
\end{align}
The last line is the contribution of the light and strange quarks to
the flow. The fermion loop yields the negative sign and the particle,
antiparticle with spin $1/2$ the factor four. The other lines are the
contributions from the mesons. Owing to light isospin symmetry, the
respective masses of the $a_0$-, $\kappa$-, $\pi$- and $K$-mesons
degenerate and their contributions can be subsumed correspondingly
into one threshold function with the respective multiplicity. The
bosonic and fermionic threshold functions $l_0^{(B,F)}$ are functions
of the dimensionless masses $\bar m_{\Phi,k} = M_{\Phi,k}/k$, the
anomalous dimension of the corresponding field as well as temperature
and chemical potential. They are defined in App.~\ref{app:rucou},
Eqs.~\eq{eq:l0b} and \eq{eq:l0f}. We cast this partial differential
equation into a finite set of coupled ordinary differential equations
by using the two-dimensional fixed background Taylor expansion, see
App.~\ref{app:taylor}.

The explicit chiral symmetry breaking terms in \Eq{eq:effpot} receive
a canonical running after the rescaling with the wave function
renormalizations discussed above. We immediately infer from
$\bar j_{L/S} = j_{L/S}/Z_{\phi,k}^{1/2}$ and \Eq{eq:etadef} that
\begin{align}
\partial_t \bar j_{L/S} = \frac{1}{2} \eta_{\phi,k} \bar j_{L/S}\,.
\end{align}
Hence, even though explicit chiral symmetry breaking is introduced by
constant source terms, reparametrization invariance results in a
running of these sources.

In the case of the explicit axial symmetry breaking parameter $c_k$,
we have to be a bit more careful. Since such a term arises originally
from instanton contributions \cite{'tHooft:1976up}, it depends in
general on the scale as well as temperature and chemical
potential. Determining the precise running of the axial anomaly is
beyond the scope of the present work. We rather resort to an effective
resolution of this issue. While the fate of the axial anomaly at
temperatures close to $T_c$ is still under debate \cite{Cohen:1996ng,
  Birse:1996dx, Kapusta:1995ww, Mitter:2013fxa}, recent results from
lattice gauge theory studies indicate that the axial symmetry is not
restored for temperatures $T\lesssim 200$ MeV \cite{Bazavov:2012qja,
  Buchoff:2013nra, Sharma:2013nva, Bhattacharya:2014ara,
  Dick:2015twa}. We therefore assume a constant KMT coupling
$\bar c = c_k / Z_{\phi,k}^{3/2}$, i.e.,
\begin{align}
\partial_t \bar c = 0\,.
\end{align}
This is equivalent to
$\partial_t c_k = -\frac{3}{2}\eta_{\phi,k} c_k$. On the other hand, if we would
require $\partial_t c_k = 0$, the running anomaly strength $\bar c$
would increase with increasing temperature, in particular for
$T \gtrsim T_c$. As we will see later this is because of
the rapid decrease of the meson anomalous dimension
with temperature in this region, cf. Fig.~\ref{fig:z}. However,
$\bar c$ should decrease at high temperatures since instanton
fluctuations are screened by the Debye mass \cite{Gross:1980br}.  Due
to the finite UV cutoff the temperature independency of the initial
action is also limited. For our UV cutoff choice the temperature
should not exceed $T\approx 200\,\text{MeV}$ as discussed in
Sec.~\ref{sec:ini}.  Hence, a constant $\bar c$ is a reasonable
assumption which is also in accordance with numerical lattice
simulations.

%%%%%%%%%%%%%%%%%%%%
\subsection{Running Yukawa Couplings}\label{sec:yuk}

%%
%%%%%%%%%%%%%%%
\begin{figure}[t]
\begin{center}
 \includegraphics[width=.82\columnwidth]{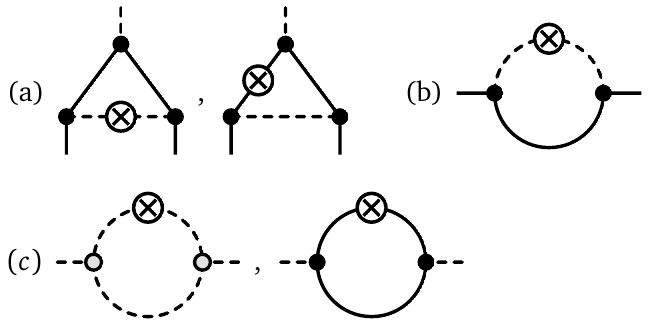}
\end{center}
\caption{The different types of diagrams that contribute to the
  running of the Yukawa couplings (a), the quark anomalous dimensions,
  (b), and the meson anomalous dimensions, (c). The solid lines are
  the quark propagators and the dashed lines are the meson
  propagators. The crossed circle illustrates the insertion of the
  scale derivative of the regulator. The black dots denote the Yukawa
  couplings and the gray dots the three-meson interactions. Note that
  the vertices and propagators are fully dressed here and the diagram
  with a quark regulator insertion vanishes for the anomalous
  dimension.}\label{fig:dcont}
\end{figure}
%%%%%%%%%%%%%%%
%%

The correlations between quarks, antiquarks and mesons, i.e.,
correlators of the form $\langle \bar q \phi q \rangle$, are
incorporated by the Yukawa couplings $h_{q,k}$ given in
\Eq{eq:zandh}. The most relevant correlation functions of this kind
are those that directly contribute to the spontaneously generated
constituent quark masses, \Eq{eq:qmasses}. We will therefore
concentrate on the running of the diagonal entries $h_{l,k}$ and
$h_{s,k}$.  This leads to a closed set of flow equations for our
truncation \Eq{eq:trunc} if we also restrict ourselves to meson
anomalous dimensions of purely light and strange mesons.  Hence, the
remaining off-diagonal Yukawa couplings in \Eq{eq:zandh} do not enter
the system of flow equations in this case.

In Ref.~\cite{Pawlowski:2014zaa} it was shown that the flow of the
Yukawa couplings can be extracted from the flow of the quark two-point
function since they are directly related to the running of the quark
masses
\begin{align}\label{eq:hproj}
\begin{split}
  \partial_t h_{l,k} &= \frac{-i}{4 N_c \sigma_L} \lim_{p\rightarrow
    p_\text{ext}}\text{Tr}\left(\frac{\delta^2\partial_t\Gamma_k}{\delta
      l(-p) \delta \bar l(p)}\right)\bigg|_{\Phi_0}\,,\\ 
\partial_t h_{s,k} &= \frac{-i \sqrt{2}}{4 N_c \sigma_S}
\lim_{p\rightarrow
  p_\text{ext}}\text{Tr}\left(\frac{\delta^2\partial_t\Gamma_k}{\delta
    s(-p) \delta \bar s(p)}\right)\bigg|_{\Phi_0}\,. 
\end{split}
\end{align}
Because the Yukawa couplings are correlation functions that carry a
baryon number some caution is advised in order to fulfill the Silver
Blaze property of QCD \cite{Cohen:2003kd}.  It states that at
vanishing temperature, the QCD equation of state does not depend on
the chemical potential as long as it is below a critical one. This
directly converts to the frequency dependence of $n$-point functions
with legs that carry a non-vanishing baryon number $B$. In the context
of the FRG, this is discussed in detail in \cite{Khan:2015puu} and
applied in \cite{Fu:2015naa, Fu:2016tey}. It follows that the Silver
Blaze property is respected if one evaluates the flow of the
correlation functions at complex frequencies
$p_0 = \tilde p_0 - 3 i B\mu$, where $\tilde p_0$ is an arbitrary real
number.  As a consequence the external four-momentum $p_\text{ext}$ at
which we evaluate correlation functions that involve quarks with
baryon number $B=1/3$ are
\begin{align}\label{eq:pext}
p_\text{ext} = (p_{0,\text{ext}},\vec{p}_\text{ext}) = (\pi T - i \mu, \vec{0})\,.
\end{align}
We have chosen the lowest fermionic Matsubara frequency for
$\tilde p_{0,\text{ext}} = \pi T$ since quark legs in loop diagrams
carry fermionic momenta. The spatial external momentum is set to zero
for convenience. A diagrammatic representation of the typical
contributions to the flow of the Yukawa couplings is shown in
Fig.~\ref{fig:dcont}.

Inserting $\Gamma_k$ from \Eq{eq:trunc} into \Eq{eq:hproj} we obtain
with the definition (\ref{eq:ruh}) the flows of the running Yukawa
couplings
\begin{align}\label{eq:hlflow}
\partial_t \bar h_{l,k} &= \left( \eta_{l,k}+\frac{1}{2}\eta_{\phi,k} \right) \bar h_{l,k} + \frac{\bar h_{l,k}^3}{4 \pi^2}\,\bigg(\! \cos^2\!\varphi_s\, L_{(1,1)}^{(l,f_0)}\\\nonumber
&\quad + L_{(1,1)}^{(l,a_0)} + \sin^2\!\varphi_s\, L_{(1,1)}^{(l,\sigma)}-\cos^2\!\varphi_p\, L_{(1,1)}^{(l,\eta)} - L_{(1,1)}^{(l,\pi)}\\\nonumber
&\quad - \sin^2\!\varphi_p\, L_{(1,1)}^{(l,\eta^\prime)} \bigg)\,,
\end{align}
for the light quark and
\begin{align}\label{eq:hsflow}
\partial_t \bar h_{s,k} &= \left( \eta_{s,k}+\frac{1}{2}\eta_{\phi,k} \right) \bar h_{s,k} + \frac{\bar h_{s,k}^3}{4 \pi^2}\,\bigg(\! \sin^2\!\varphi_s\, L_{(1,1)}^{(s,f_0)}\\\nonumber
&\quad + \cos^2\!\varphi_s\, L_{(1,1)}^{(s,\sigma)} -\sin^2\!\varphi_p\, L_{(1,1)}^{(s,\eta)}  - \cos^2\!\varphi_p\, L_{(1,1)}^{(s,\eta^\prime)} \bigg)\,,
\end{align}
for the strange quark Yukawa coupling.  The threshold functions for
loops with internal quarks and mesons have been abbreviated by
$L_{(1,1)}^{(q,\phi)}=L_{(1,1)}(\bar m_{q,k}^2,\bar
m_{\phi,k}^2,\eta_{q,k},\eta_{\phi,k};p_{0,\text{ext}})$
and are defined in App.~\ref{app:rucou}, \Eq{eq:l11}.

Due to the mixing of the physical $f_0,\sigma$ and $\eta,\eta^\prime$
states with the corresponding purely light and strange states as
defined in \Eq{eq:plstrafo}, the scalar and pseudoscalar mixing angles
$\varphi_s,\varphi_p$ naturally appear here.  These equations couple
to the flow of the effective potential through the quark masses.

%%%%%%%%%%%%%%%%%%%%
\subsection{Anomalous Dimensions}\label{sec:etas}

The wave function renormalizations encode the fluctuation-induced
modifications of the corresponding fields or in other words, the
deviation from the canonical momentum dependence of the
propagators. In the present context, this is particularly important
for the identification of Euclidean meson masses. The physical masses
are defined as poles of the corresponding propagators or as roots from
the inverse propagator $\bar\Gamma^{(2)}(p_0,\vec{p}^2)$ through
$\bar\Gamma^{(2)}(i m_{\text{pol}},0) = 0$. In this work, we instead
extract the meson masses from the curvature of the effective potential
given by $\bar\Gamma^{(2)}(0,0) = m_{\text{cur}}^2$ evaluated at
vanishing momenta. The computation of the pole masses would require
the analytical continuation of the Euclidean two-point functions. Even
though it is known how this can be achieved within the FRG
\cite{Floerchinger:2011sc, Tripolt:2013jra, Pawlowski:2015mia}, it is
beyond the scope of the present work.  However, for an effective
theory setup within the FRG the difference between the pole
$m_{\text{pol}}$ and curvature $m_{\text{cur}}$ masses should remain
small as pointed out in \cite{Strodthoff:2011tz}. This matters because
some of the physical meson masses are used as input to constrain some
parameters of our effective theory.

For the utilized truncation the inverse propagator is parametrized by
\begin{align}
\bar\Gamma^{(2)}(p_0,\vec{p}^2) = p_0^2+\vec{p}^2 + \frac{m^2}{Z(p_0,\vec{p}^2)}\,,
\end{align}
which yields the relation 
\begin{align}
m_{\text{cur}}^2 = \frac{Z(i m_{\text{pol}},0)}{Z(0,0)} m_{\text{pol}}^2\,,
\end{align}
between the curvature and pole mass.  Hence, the pole and curvature
masses are essentially identical if the wave function renormalization
only mildly depends on the (imaginary) frequency.  This mild
dependency has been confirmed in \cite{Helmboldt:2014iya} for pions in
a two-flavor study. In addition, the inclusion of the RG-scale
dependent, but momentum-independent, wave function renormalizations
$Z_k$ yields an agreement between the pole and curvature mass of the
pion within an one percent accuracy level \cite{Helmboldt:2014iya}.
Thus, by including the running wave function renormalizations, we
ensure that the computed masses stay close to the physical ones which
furthermore justifies our model parameter fixing based on measured
masses.

Another important issue that emphasizes in particular the relevance of
the meson anomalous dimension $Z_{\phi,k}$ is related to the
decoupling of the mesons from the physical spectrum in the quark-gluon
regime of QCD at large energies. As shown in \cite{Braun:2014ata,
  Rennecke:2015eba} a rapid decrease of the meson wave function
renormalization drives this decoupling, see also \cite{Gies:2002hq, Braun:2008pi}. This is also the case for
three quark flavors: even without the dominating gluon fluctuations,
$Z_{\phi,k}$ shows a rapid decrease for temperatures above the chiral
transition, $T \!>\! T_c$. The resulting fast decoupling of the mesons
in the chirally symmetric phase has a large impact on the phase
boundary and in particular on the location of the critical endpoint as
compared to previous results where this effect has been neglected
\cite{Mitter:2013fxa}.

%%
%%%%%%%%%%%%%%%
\begin{table*}
\begin{center}
\begin{tabular}{ l l l l l l l l l l }
  \hline\hline truncation &\; running couplings &\;
                                                  $\bar\lambda_{10,\Lambda}$
                                                  [GeV$^2$] &\;
                                                              $\bar\lambda_{20,\Lambda}$
  &\; $\bar\lambda_{01,\Lambda}$ &\; $\bar c$ [GeV] &\; $\bar j_L$
                                                      [GeV$^3$] &\;
                                                                  $\bar
                                                                  j_S$
                                                                  [GeV$^3$]
  &\; $\bar h_{l,k}$ &\; $\bar h_{s,k}$ \\ 
\hline\vspace{-0.4cm}\\ 
 LPA$^\prime$+Y &\; $\bar{\tilde U}_k$, $\bar h_{l,k}$, $\bar
                  h_{s,k}$, $Z_{l,k}$, $Z_{s,k}$, $Z_{\phi,k}$ &\;
                                                                 $(1.221)^2$
                                                            &\; 60 &\;
                                                                     210
                                 &\; 4.808 &\; $1.495\times(0.121)^3$
                                                                &\;
                                                                  $1.495\times(0.336)^3$
  &\; 9.78 &\; 9.65\\ 
 LPA+Y &\; $\bar{\tilde U}_k$, $\bar h_{l,k}$, $\bar h_{s,k}$ &\;
                                                                $(0.836)^2$
                                                            &\; 7.5
  &\; 50 &\; 4.808 &\; $1\times(0.121)^3$ &\; $1\times(0.336)^3$ &\;
                                                                   5.96
                     &\; 6.40\\ 
 LPA &\; $\bar{\tilde U}_k$ &\; $(0.889)^2$ &\; 4.5 &\; 40 &\; 4.808
                                                    &\;
                                                      $1\times(0.121)^3$
                                                                &\;
                                                                  $1\times(0.336)^3$
  &\; 6.5 &\; 6.5\\ 
  \hline\hline
\end{tabular}
\end{center}
\caption{Initial conditions for different truncations.}
\label{tab:init}
\end{table*}
%%%%%%%%%%%%%%%
%%

The running of the wave function renormalizations gives rise to the
anomalous dimension defined in \Eq{eq:etadef}. For the light and
strange quarks, they can be obtained from the corresponding two-point
functions by
\begin{align}
\begin{split}
\eta_{l,k} &= - \frac{1}{16 N_c Z_{l,k}} \lim_{p\rightarrow
  p_\text{ext}} \frac{\partial^2}{\partial |\vec{p}|^2}
\text{Tr}\left(\!\vec{\gamma}\vec{p}\,\frac{\delta^2\partial_t\Gamma_k}{\delta
    l(-p) \delta \bar l(p)}\right)\bigg|_{\Phi_0}\,,\\ 
\eta_{s,k} &= - \frac{1}{8 N_c Z_{l,k}} \lim_{p\rightarrow
  p_\text{ext}} \frac{\partial^2}{\partial |\vec{p}|^2}
\text{Tr}\left(\!\vec{\gamma}\vec{p}\,\frac{\delta^2\partial_t\Gamma_k}{\delta
    s(-p) \delta \bar s(p)}\right)\bigg|_{\Phi_0}\,. 
\end{split}
\end{align}
Concerning the Silver Blaze property at vanishing temperature the same
arguments as for the Yukawa coupling apply and we thus evaluate these
flows at external momenta $p_\text{ext}$ as given in \Eq{eq:pext}
which finally yields
\begin{align}\label{eq:etal}
\begin{split}
\eta_{l,k} &= \frac{\bar h_{l,k}^2}{24 \pi^2} (4-\eta_{\phi,k})\, \bigg(\! \cos^2\!\varphi_s\, \textit{FB}_{(1,2)}^{(l,f_0)} + \textit{FB}_{(1,2)}^{(l,a_0)}\\
&\quad+ \sin^2\!\varphi_s\,\textit{FB}_{(1,2)}^{(l,\sigma)} + \cos^2\!\varphi_p\, \textit{FB}_{(1,2)}^{(l,\eta)} + \textit{FB}_{(1,2)}^{(l,\pi)}\\
&\quad+ \sin^2\!\varphi_p\, \textit{FB}_{(1,2)}^{(l,\eta^\prime)} \bigg)
\end{split}
\end{align}
for the light and
\begin{align}\label{eq:etas} \nonumber
\eta_{s,k} &= \frac{\bar h_{s,k}^2}{24 \pi^2} (4-\eta_{\phi,k})\, \bigg(\! \sin^2\!\varphi_s\, \textit{FB}_{(1,2)}^{(s,f_0)} + \cos^2\!\varphi_s\,\textit{FB}_{(1,2)}^{(s,\sigma)}\\
&\quad + \sin^2\!\varphi_p\, \textit{FB}_{(1,2)}^{(s,\eta)}+ \cos^2\!\varphi_p\, \textit{FB}_{(1,2)}^{(s,\eta^\prime)} \bigg)
\end{align}
for the strange quark anomalous dimension. For the threshold functions
we have used the abbreviation
$\textit{FB}_{(1,2)}^{(q,\phi)} = \textit{FB}_{(1,2)}(\bar
m_{q,k}^2,\bar
m_{\phi,k}^2,\eta_{q,k},\eta_{\phi,k};p_{0,\text{ext}})$,
see App.~\ref{app:rucou}, Eqs.~\eq{eq:fb11} and \eq{eq:fbnm} for the
definitions. The type of diagram that contributes to these anomalous
dimensions is displayed in Fig.~\ref{fig:dcont}.

In Eqs.~\eq{eq:etal} and \eq{eq:etas} our assumption of only one
anomalous dimension $\eta_{\phi,k}$ simplifies the flow equations
significantly since otherwise each meson term would have its own
prefactor $(4-\eta_{\phi_i,k})$.  Of course, which of the anomalous
dimension as above \Eq{eq:1} is chosen is not unique.  Since the light
pions are dynamically the most relevant degrees of freedom their
dynamics should be captured as accurately as possible.  This motivates
our choice
\begin{align}\label{eq:zphidef}
Z_{\phi,k} \equiv Z_{\pi^+,k}\,,
\end{align}
where actually the pion charge is irrelevant due to the light isospin
symmetry.  The light isospin symmetry also implies the vanishing of
the Yukawa couplings for the $\pi^0$ and for the $a_0^0$.  However,
this approximation could potentially lead to an overestimation of the
dynamics of the heavier mesons at large temperatures.  A reasonable
upgrade of the present approximation would be the inclusion of further
meson wave function renormalizations, e.g., a second meson wave
function renormalization of a strange meson. This might be of
relevance for further investigations of cumulants for strangeness
number distribution.  However, we will defer such an analysis to a
future work.

The meson anomalous dimension defined in \Eq{eq:zphidef} can be
extracted from the pion two-point function by
\begin{align}
\eta_{\phi,k} = -\frac{1}{2 Z_{\phi,k}} \lim_{p\rightarrow0}
  \frac{\partial^2}{\partial |\vec{p}|^2} \text{Tr}
  \left(\frac{\delta^2 \partial_t \Gamma_k}{\delta
  \pi^+(-p)\delta\pi^+(p)}\right)\bigg|_{\Phi_0}\,. 
\end{align}
Here the Silver Blaze property is not an issue since the anomalous
dimension is related to a purely mesonic correlator.  We therefore set
the external frequency to the lowest bosonic Matsubara mode and the
external spatial momenta to zero, i.e.,
$p_{\text{ext}} =(0, \vec{0})$. This yields 
\begin{align}\label{eq:etaphi}
\begin{split}
\eta_{\phi,k} &= \frac{1}{3 \pi^2} \bigg[
\bar\lambda_{\pi^+\pi^-f_0,k} \textit{BB}_{(2,2)}^{(\pi,f_0)} +
\bar\lambda_{\pi^+\pi^-\sigma,k} \textit{BB}_{(2,2)}^{(\pi,\sigma)}\\ 
&\quad \bar\lambda_{\pi^+a_0^-\eta,k} \textit{BB}_{(2,2)}^{(a_0,\eta)}
+ \bar\lambda_{\pi^+a_0^-\eta^\prime,k}
\textit{BB}_{(2,2)}^{(a_0,\eta^\prime)}\\ 
&\quad+\big(\bar\lambda_{\pi^+\kappa^-\bar
  K^0,k}+\bar\lambda_{\pi^+K^-\bar \kappa^0,k}\big)
\textit{BB}_{(2,2)}^{(K,\kappa)}\bigg]\\ 
&\quad + \frac{N_c\, \bar h_{l,k}^2}{6 \pi^2} \bigg[
4\big(2-\eta_{l,k}\big) \textit{F}_{(3)}^{\,(l)} -
\big(3-2\eta_{l,k}\big) \textit{F}_{(2)}^{\,(l)}\bigg]\,. 
\end{split}
\end{align}
The involved meson three-point correlators
$\bar\lambda_{\phi_i\phi_j\phi_l,k}$ of the mesons $\phi_i$ are given
in App.~\ref{app:mesmc}. Light isospin symmetry implies
$\bar\lambda_{\pi^+\kappa^-\bar K^0,k} = \bar\lambda_{\pi^+K^-\bar
  \kappa^0,k}$.
The threshold functions
$\textit{BB}_{(2,2)}^{(\phi_i,\phi_j)} = \textit{BB}_{(2,2)}(\bar
m_{\phi_i,k}^2, \bar m_{\phi_j,k}^2)$
are related to purely mesonic loops but for different mesons and
$\textit{F}_{(2)}^{\,(l)} = \textit{F}_{(2)}(\bar m_{l,k}^2)$
corresponds to a light quark loop.  The analytical expressions for
these functions are given in App.~\ref{app:rucou}. The corresponding
diagrams that contribute to the meson anomalous dimension are again
shown in Fig.~\ref{fig:dcont}.

%%%%%%%%%%%%%%%%%%%%%%%%%%%%%%%%%%%%%%%%%%%%%%%%%%%%%%%%%%%%%%
%%%%%%%%%%%%%%%%%%%%%%%%%%%%%%%%%%%%%%%%%%%%%%%%%%%%%%%%%%%%%%
\section{Results}\label{sec:res}

\subsection{Initial Conditions}\label{sec:ini}

We begin with the specification of the initial effective action
$\Gamma_{k=\Lambda}$ at the UV scale $\Lambda$.  Within our low-energy
description the initial scale $\Lambda$ is bounded on the one hand by
the requirement that is has to be large enough to avoid UV-cutoff
effects at large temperature and/or chemical potential and on the
other hand it should be as small as possible to ensure a decoupled
gauge sector. Satisfying both of these requirements is hardly possible
since the gluon sector remains quantitatively relevant even at scales
below the chiral transition \cite{Rennecke:PhD}.  We therefore choose
as smallest UV-cutoff $\Lambda = 900$ MeV for which cutoff effects are
only minor for $T\lesssim 230\,\text{MeV}$ and
$\mu\lesssim 300\,\text{MeV}$.

Furthermore, the initial effective potential
$\bar{\tilde U}_{k=\Lambda}$ and the initial Yukawa couplings
$\bar h_{q,k=\Lambda}$ have to be specified.  For the wave function
renormalizations no initial conditions have to be given explicitly
since due to the reparametrization invariance of the RG-equations only
the anomalous dimensions enter the flow equations. For the initial
effective potential only relevant and marginal terms are chosen
\begin{align}
\begin{split}
\bar{\tilde U}_{k=\Lambda}(\bar\rho_1,\bar{\tilde\rho}_2) &=
\bar\lambda_{10,k=\Lambda}\, \bar\rho_1 +
\frac{\bar\lambda_{20,k=\Lambda}}{2}\, \bar\rho_1
+\bar\lambda_{01,k=\Lambda}\, \bar{\tilde\rho}_2\\ 
&\quad - \bar c \bar\xi -\bar j_L \bar\sigma_L - \bar j_S 
\bar\sigma_S\,, 
\end{split}
\end{align}
since higher-order meson correlations are generated via the flow at
lower energy scales.

Together with the initial Yukawa couplings we have eight parameters to
fix in the infrared. As input we use the pion and kaon decay
constants, $\bar f_\pi = 93\,\text{MeV}$ and $\bar f_K = 114\,\text{MeV}$, the
curvature masses of the pion, $M_{\pi} = 138\,\text{MeV}$, the kaon,
$M_{K} = 492\,\text{MeV}$, the sigma meson,
$M_{\sigma} = 421\,\text{MeV}$ and the sum of the squared $\eta$- and
$\eta^\prime$-masses
$M_{\eta}^2 + M_{\eta^\prime}^2 = 1.22 \times 10^6\,\text{MeV}^2$.
Furthermore, the constituent quark masses are fixed to
$M_{l} = 301\,\text{MeV}$ and $M_{s} = 441\,\text{MeV}$.

However, these infrared conditions do not fix all initial values
uniquely in the UV. This additional ambiguity can be utilized to
imprint some information of the full RG flow of QCD onto our
truncation.  Since $\Lambda$ is chosen to be much larger than the
chiral symmetry breaking scale, mesons should be already decoupled
from the system and act merely as auxiliary fields.  This was
explicitly demonstrated for $N_f=2$ QCD in \cite{Braun:2014ata,
  Rennecke:2015eba}. We implement this effectively by minimizing the
meson fluctuations at large cutoff scales and require their initial
masses to be larger than the cutoff scale, i.e.,
$M_{\phi,k=\Lambda} > \Lambda$.  Since the computed curvature masses
are expected to be close to the pole masses, as discussed in
Sec.~\ref{sec:etas}, the condition $M_{\phi,k=\Lambda} > \Lambda$
implies that the mesons do not emerge as resonances in the
corresponding spectral functions and are indeed decoupled from the
physical spectrum.

To point out the physical effects of the applied effective model, we
solve the flow equation system in different truncations. To fix the
terminology, we denote the truncation of the effective action we
introduced in Sec.~\ref{sec:flucts} as "LPA$^\prime$+Y", i.e., we
consider the full effective potential $\bar{\tilde U}_k$ together with
the running Yukawa couplings discussed in Sec.~\ref{sec:yuk} as well
as non-vanishing anomalous dimension as discussed in
Sec.~\ref{sec:etas}. Most commonly used in the literature is the local
potential approximation, "LPA", where only the scale-dependent
effective potential is investigated. It can by obtained from the
LPA$^\prime$+Y schema by considering constant Yukawa couplings,
$\partial_t \bar h_{q,k} = 0$, and setting the anomalous dimensions to
zero, $\eta_{\Phi,k} = 0$.  A third simplified schema refines the LPA
truncation by allowing for running Yukawa couplings which we label by
"LPA+Y".  It includes the running of the Yukawa couplings
$\bar h_{q,k}$ but ignores the anomalous dimensions. The initial
conditions for the different truncations are summarized in
Tab.~\ref{tab:init}.

%%
%%%%%%%%%%%%%
\begin{figure}[t]
\begin{center}
  \includegraphics[width=1.\columnwidth]{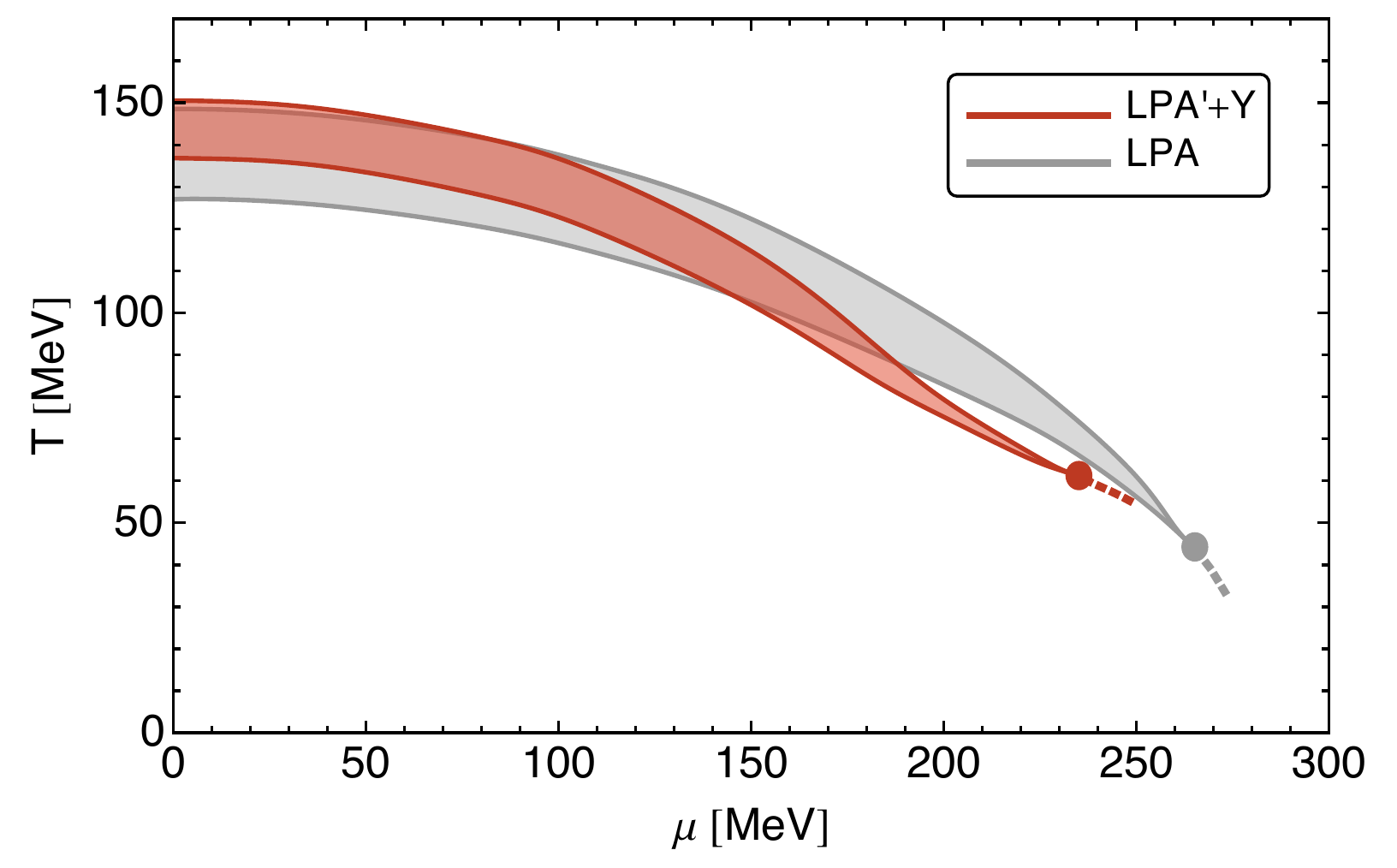}
  \caption{The phase boundaries of the chiral transition in the LPA'+Y
    and LPA truncations. The crossover transitions corresponds to the
    width of the light $\partial \bar\sigma_{L,k=0}/\partial T$ at
    80\% of its peak height. The dots mark the critical endpoints and
    the dotted lines the first-order transitions.}\label{fig:pb}
\end{center}
\end{figure}
%%%%%%%%%%%%%
%%

\subsection{The Phase Boundary and the Critical Endpoint}
\label{sec:pb}

The result for the chiral phase boundary is shown in Fig.~\ref{fig:pb}
wherein the red (dark) region has been obtained with the LPA$^\prime$+Y
scheme. Since the transition is a crossover at small $\mu$, we show
the width of the light condensate in the IR,
$\partial \bar\sigma_{L,k=0}/\partial T$ at 80\% of its peak height.
The maximum at $\mu=0$ corresponds to a pseudocritical
$T_c = 144\,\text{MeV}$.  The width stays almost constant until
$\mu\sim 150\,\text{MeV}$ and then the transition becomes sharper for
larger $\mu$. At $\mu = 235\,\text{MeV}$ the width is zero and the
transition is of second order. For $\mu > 235\,\text{MeV}$ we find a
first-order transition which is indicated by the dotted red line.
Since the expansion of the effective potential is not fully converged
in this region we stop the computation for $\mu>250\,\text{MeV}$. For
details on the expansion we refer to App.~\ref{app:taylor}.

To emphasize the physical effects caused by the running Yukawa
couplings and wave function renormalizations, we compare the
LPA$^{\prime}$+Y results to the ones obtained in the LPA truncation,
where these contributions are neglected. The corresponding phase
boundary is given by the gray (light) band in Fig.~\ref{fig:pb}. The
width of the phase boundary at vanishing chemical potential is about
50\% larger in LPA as compared to the LPA$^\prime$+Y truncation. Thus,
the overall effect of the running Yukawa couplings and the anomalous
dimensions is to sharpen the phase transition.

\begin{figure}[t]
\begin{center}
  \includegraphics[width=.96\columnwidth]{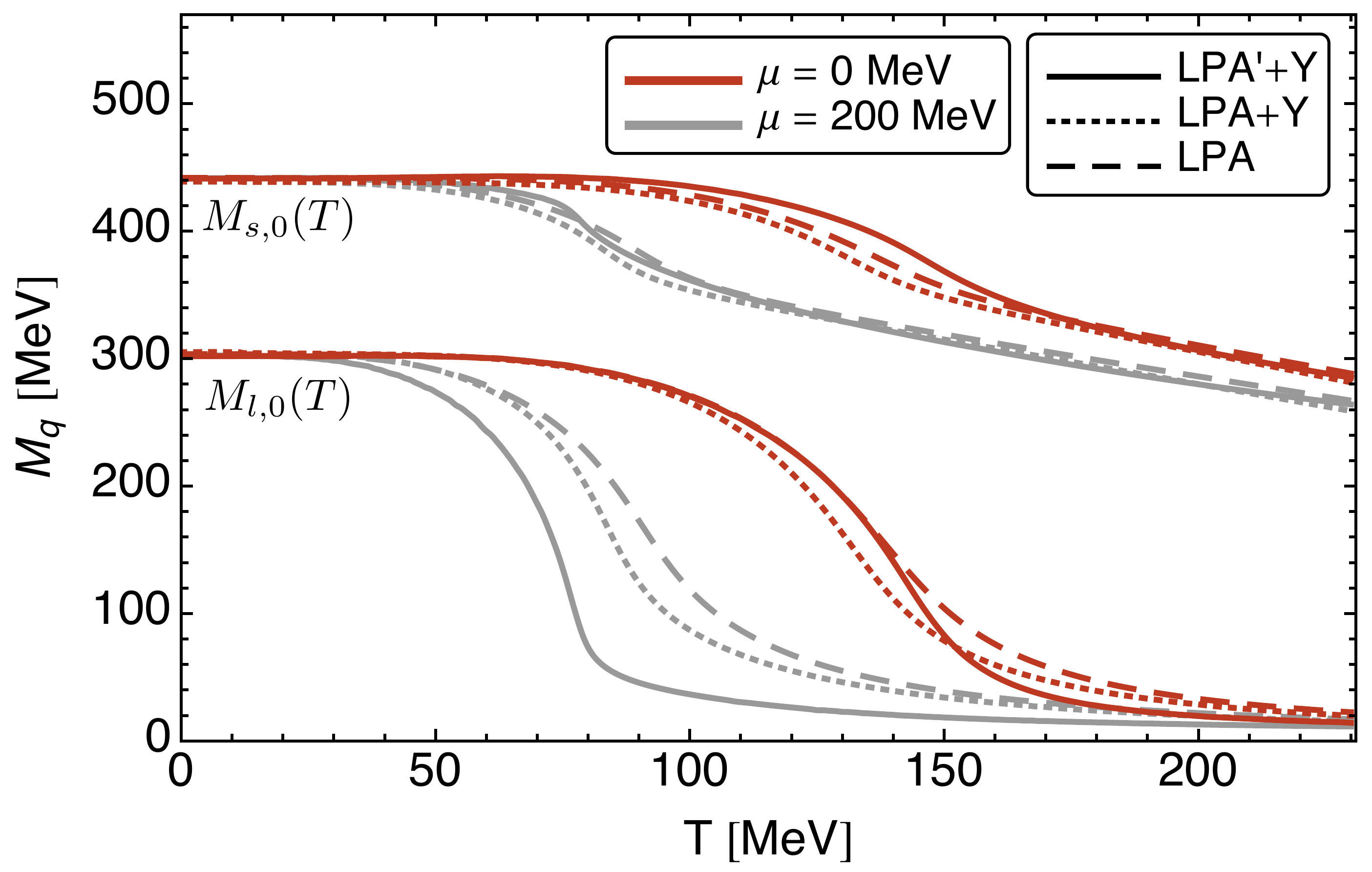}
  \caption{The light and strange quark masses in the IR as a function
    of temperature for the three truncations at $\mu=0$ and
    $\mu=200\,\text{MeV}$.}\label{fig:mq}
\end{center}
\end{figure}

In order to distinguish between the various contributions of the
Yukawa couplings and of the anomalous dimensions, we show in
Tab.~\ref{tab:cep} the location of the critical endpoint in dependence
of the three truncation schema.
%%
%%%%%%%%%%%%%
\begin{table}[h]
\begin{center}
\begin{tabular}{ l l }
  \hline\hline truncation &\quad $(T_\text{CEP}, \mu_\text{CEP})$ [MeV] \\
\hline
 LPA$^\prime$+Y &\quad (61, 235) \\
 LPA+Y &\quad (46, 255) \\
 LPA &\quad (44, 265) \\
  \hline\hline
\end{tabular}
\end{center}
\caption{Location of the critical endpoint in various truncations.}
\label{tab:cep}
\end{table}
%%%%%%%%%%%%%
%%
In LPA it is located at $(T_\text{CEP},\mu_\text{CEP})=(44, 265)$ MeV
and in LPA$^{\prime}$+Y at $(T_\text{CEP},\mu_\text{CEP})=(61, 235)$
MeV.  While both the running Yukawa couplings as well as the anomalous
dimensions push the CEP to smaller chemical potentials and larger
temperatures, the anomalous dimensions yield the dominant
contribution. The effects of the different truncations also become
apparent in the quark masses which are presented in
Fig.~\ref{fig:mq}. Both, the Yukawa coulings and the wave function
renormalizations sharpen the chiral transition. This effect is more
pronounced at larger chemical potential. We conclude that the
fluctuation-induced corrections to the quark and meson propagators,
i.e., the wave function renormalizations, as well as quantum
corrections to the quark-meson scattering processes, i.e., the running
Yukawa couplings, have a substantial quantitative effect on the chiral
phase boundary.

%%
%%%%%%%%%%%%%
%%%%%%%%%%%%%
%%

\begin{figure}[t]
\begin{center}
  \includegraphics[width=.98\columnwidth]{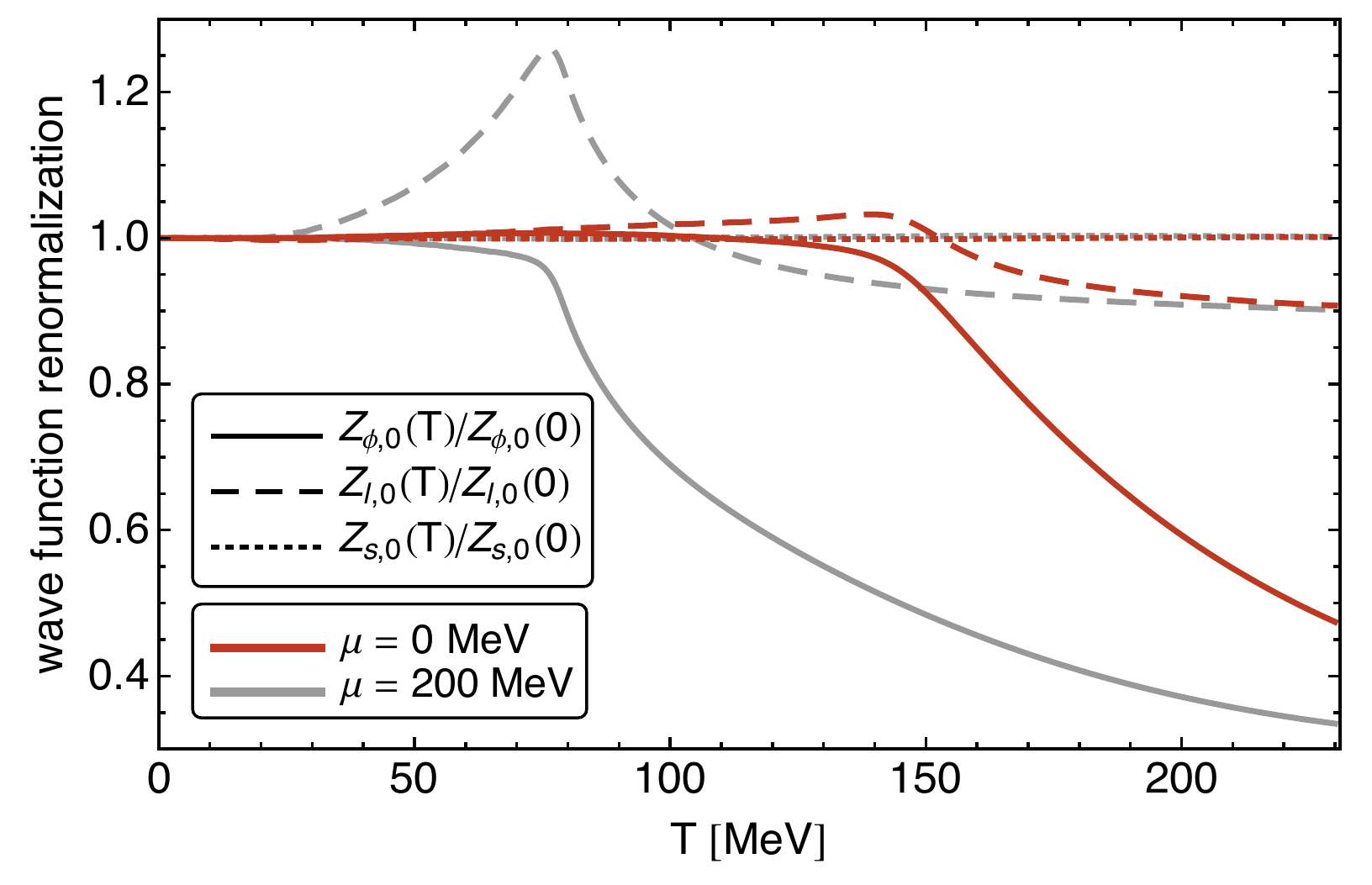}
  \caption{Meson, light and strange quark wave function
    renormalizations as a function of the temperature at $\mu=0$ and
    $\mu=200$ MeV. The strange $Z_{s,k=0}$ is almost temperature
    independent since it varies only about $0.4$\% in the shown
    temperature region.}\label{fig:z}
\end{center}
\end{figure}

In order to deepen these observations, we investigate the temperature
and chemical potential dependence of the Yukawa couplings and the wave
function renormalizations in the LPA$^\prime$+Y schema. In
Fig.~\ref{fig:z} we show the temperature dependence of the meson
$Z_{\phi,k=0}$ (solid lines) as well as the light quark $Z_{l,k=0}$
(dashed lines) and strange quark $Z_{s,k=0}$ (dotted lines) wave
function renormalizations for $\mu=0$ (red dark lines) and
$\mu=200\,\text{MeV}$ (gray light lines). They are normalized to one
at vanishing temperature.  As similar to two-flavor QCD studies
\cite{Pawlowski:2014zaa, Braun:2014ata, Rennecke:2015eba}, the meson
wave function renormalizations show a characteristic behavior for
fields that eventually decouple from the physical spectrum. While the
mesonic $Z_{\phi,k=0}$ below $T_c$ is almost temperature independent,
it rapidly drops in the chirally symmetric phase. Since the wave
function renormalizations are the prefactors in the kinetic field
terms, cf.  \Eq{eq:trunc}, this behavior signals a suppression of the
mesonic fluctuations when chiral symmetry is restored.  This nicely
explains why the phase transition becomes steeper in this truncation:
The non-trivial behavior of $Z_{\phi,k=0}$ triggers a rapid decoupling
of the mesons above $T_c$. The symmetry restoring bosonic fluctuations
that tend to wash-out the phase transition are therefore suppressed in
the vicinity of the phase transition, which in turn leads to a sharper
crossover transition. This tendency becomes even more pronounced at
larger chemical potential as demonstrated by the solid gray line in
Fig.~\ref{fig:z}: the drop-off of $Z_{\phi,k=0}$ becomes steeper with
increasing $\mu$. Hence, the suppression of meson fluctuations is
amplified at larger densities which also explains qualitatively why
the CEP moves to smaller densities and larger temperatures.

%%
%%%%%%%%%%%%%
%%%%%%%%%%%%%
%%

The quark wave function renormalizations show a less prominent
behavior in Fig.~\ref{fig:z}.  In particular, the strange quark wave
function renormalization $Z_{s,k=0}$ (dashed line in the figure)
barely runs and is almost insensitive to thermal and density
fluctuations. The immediate consequence is that $Z_{s,k}$ could safely
be ignored here.  The light quark wave function renormalization
$Z_{l,k=0}$ (dashed line in the figure) first grows with increasing
temperature, before it decreases again above $T_c$. The higher the
chemical potential, the higher the peak of $Z_{l,k=0}$ at
$T_c$. Hence, light quark fluctuations are amplified through the
running light quark wave function renormalizations in the vicinity of
the phase transition. This effect becomes stronger with increasing
$\mu$. Since fermionic fluctuations have the tendency to sharpen the
chiral phase transition, the effect of the amplified quark
fluctuations adds to the effect of the meson wave function
renormalization as discussed above.

\begin{figure}[t]
\begin{center}
  \includegraphics[width=.98\columnwidth]{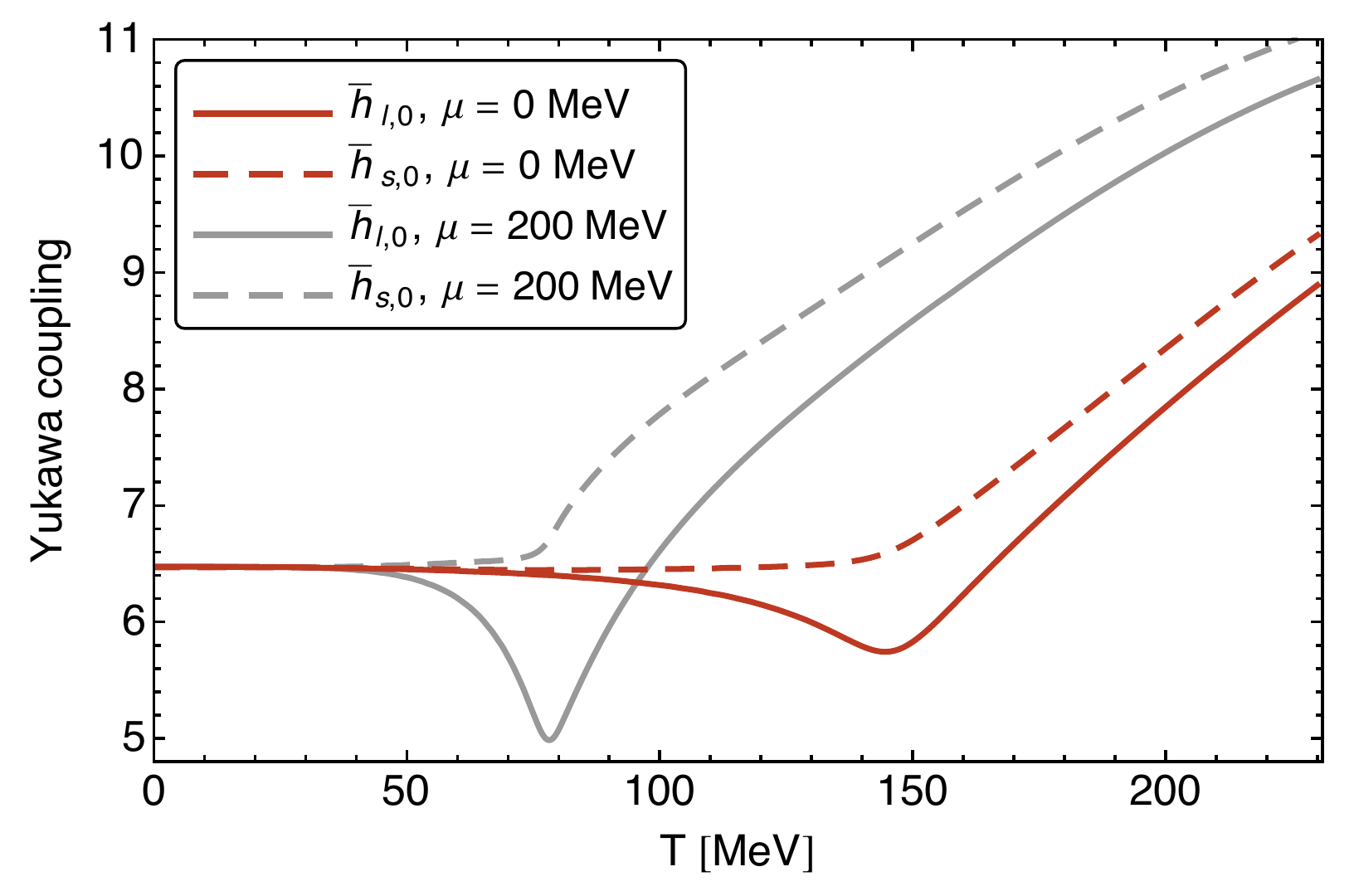}
  \caption{Light and strange quark Yukawa couplings $\bar h_{l,0}$ and
    $\bar h_{s,0}$ as a function of the temperature at $\mu=0$ and
    $\mu=200$ MeV.}\label{fig:yuk}
\end{center}
\end{figure}

In Fig.~\ref{fig:yuk} the temperature dependence of the light and
strange quark Yukawa couplings, $\bar h_{l,k=0}$ (solid line) and
$\bar h_{s,k=0}$ (dashed line) is shown at $\mu=0$ (red dark lines)
and at $\mu=200\,\text{MeV}$ (gray light lines). The initial condition
for $\bar h_{l,k}$ and $\bar h_{s,k}$ are tuned such that they yield
the same infrared vacuum value. This ensures that the difference
between the constituent masses of the light and strange quarks is
solely driven by spontaneous symmetry breaking.  They start to deviate
in the vicinity of the critical temperature. There are two opposing
contributions to the running of the Yukawa couplings. On the one hand,
there is the contribution from the triangle diagrams, i.e., the terms
proportional to $\bar h_{l,k}^3$ in \Eq{eq:hlflow} and
$\bar h_{s,k}^3$ in \Eq{eq:hsflow}. They are positive and tend to
decrease the Yukawa couplings with increasing temperature. On the
other hand, both flows have contributions from the anomalous
dimensions. From \Eq{eq:ruh} and Fig.~\ref{fig:z} we conclude that
this contribution tends to increase the Yukawa couplings with
increasing temperatures. As visible in Fig.~\ref{fig:yuk}, the
contributions from the anomalous dimensions dominate and both
couplings $\bar h_{l,k}$ and $\bar h_{s,k}$ are almost constant below
$T_c$ but increase above. The dip of $\bar h_{l,k=0}$ at $T_c$ is
directly linked to the peak of $Z_{l,k=0}$ as shown in
Fig.~\ref{fig:z}.

%%
%%%%%%%%%%%%%
%%%%%%%%%%%%%
%%

%%
%%%%%%%%%%%%%
\begin{figure}[b]
\begin{center}
  \includegraphics[width=1.\columnwidth]{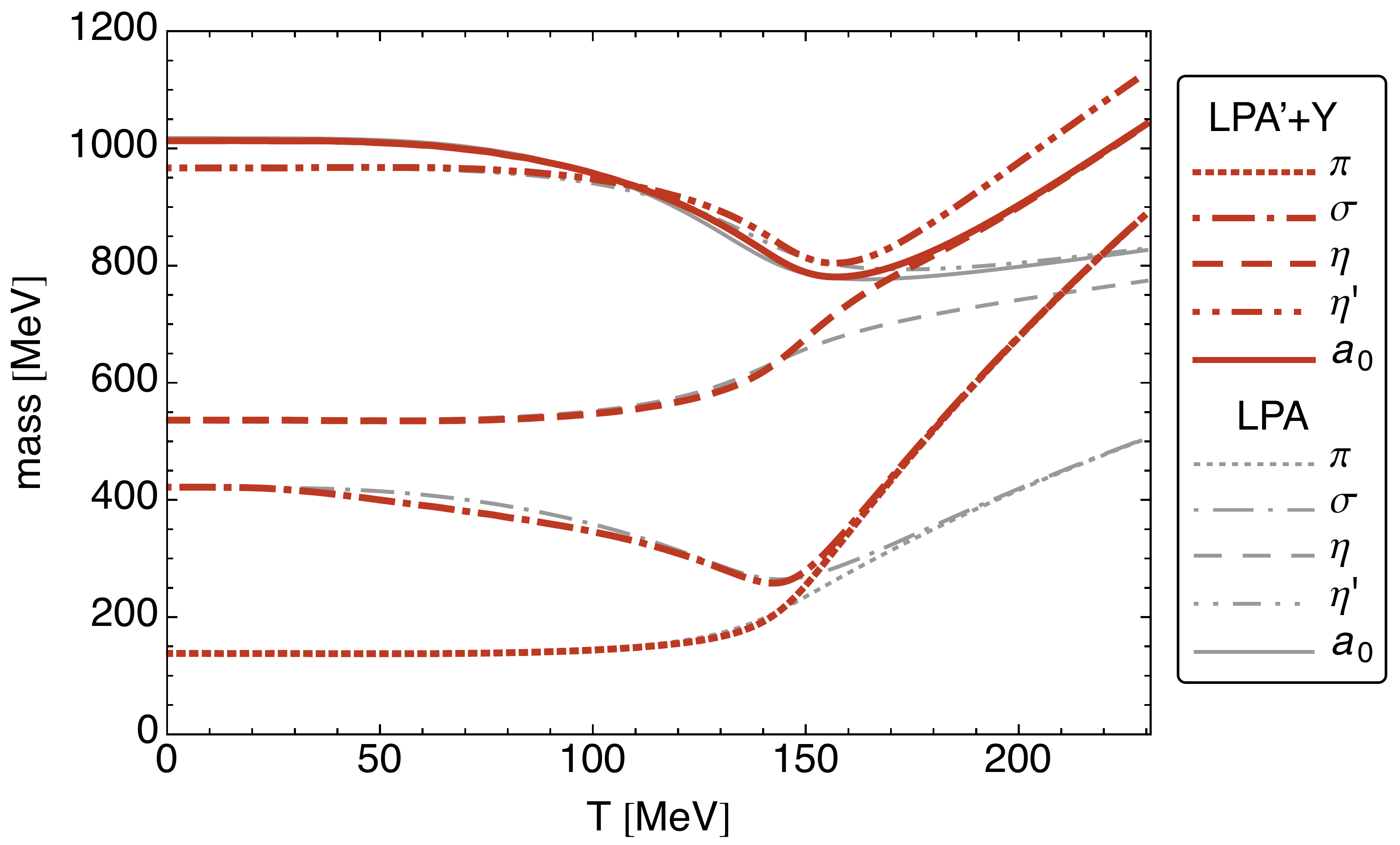}
  \caption{Meson masses as a function of the temperature at $\mu=0$ in
    LPA$^\prime$+Y (red lines) in comparison to the LPA result
    (thinner gray lines). Note, in LPA$^{\prime}$+Y the $a_0$-meson
    degenerates with the $\eta$-meson, while in LPA the $a_0$-meson
    and $\eta^\prime$-meson degenerate.}\label{fig:masses}
\end{center}
\end{figure}
%%%%%%%%%%%%%
%%

The physical effect of the meson wave function renormalizations
becomes most obvious in the meson masses as exemplified in
Fig.~\ref{fig:masses}. To demonstrate this we compare our
LPA$^{\prime}$+Y results (red dark lines) to the corresponding LPA
result (thin gray lines). As discussed above, the quantum corrections
to $Z_{\phi,k}$ drive the decoupling of the mesons in the high energy
regime which is realized by rapidly rising masses for $T>T_c$. From
Fig.~\ref{fig:masses} we conclude that the running of $Z_{\phi,k}$ is
crucial to capture this physical behavior. Interestingly, in this case
we also see a change of the chiral partner of the $a_0$-meson: While
it degenerates with the $\eta^\prime$-meson at large temperatures in
the LPA, it degenerates with the $\eta$-meson in LPA$^\prime$+Y. We
will discuss this in more detail in the next section. We also observe
a pronounced drop of the $\eta^\prime$ mass at the chiral phase
transition. This is in line with experimental results on the in-medium
modifications of this mass \cite{Csorgo:2009pa,
  Vertesi:2009wf}. However, we emphasize that this drop can be
attributed to the melting of the chiral condensates and does not
indicate the restoration of $U(1)_A$-symmetry, as also pointed out
e.g.~in \cite{Heller:2015box}. In a linear sigma model with a
temperature dependent anomaly, a drop of the $\eta^\prime$ mass is not
found \cite{Fejos:2016hbp}.

Finally, we want to discuss the systematic error of our results on the
phase boundary. Firstly and most importantly, we focused on the
effects of mesonic fluctuations on the phase boundary and neglected
the gauge sector entirely. Thus, it is impossible to deal with the
deconfinement transition in the present setup. However, it can be
upgraded in this direction in a straightforward way by effectively
including a non-vanishing temporal gluon background e.g. by means of a
Polyakov-loop enhanced quark-meson model along the lines of
\cite{Fukushima:2003fw, Ratti:2005jh, Schaefer:2007pw,
  Herbst:2010rf}. It has been shown in \cite{Herbst:2013ail} that the
inclusion of the Polyakov-loop potential tends to move the chiral
transition line as well as the CEP to higher temperatures (see
\cite{Schaefer:2009ui, Karsch:2010hm, Schaefer:2011ex} also for
similar mean-field analyses). This is reassuring since the critical
temperature at vanishing chemical potential in the present work seems
a little too low. Another shortcoming of the present truncation is the
lack of baryonic degrees of freedom. These are certainly the dominant
degrees of freedom in the high chemical potential and small
temperature regime and hence they are indispensable for a realistic
description of the QCD phase diagram in this regime, in particular for
the liquid-gas transition to nuclear matter. However, recent DSE
results indicate that baryonic degrees of freedom have only very
little effect on the chiral phase boundary \cite{Eichmann:2015kfa}.  A
direct comparison with recent DSE results on the location of the CEP
\cite{Fischer:2014ata, Fischer:2014mda} would be involved at the
present stage. For example, on the one hand gluonic effects are
neglected in this FRG study while the back-coupling of mesons on the
other hand is not explicitly taken into account in the DSE studies of
the phase diagram \cite{Fischer:2012vc}. Thus, for a sensible
comparison of the corresponding results an enlargement of the used
truncations in both functional approaches is essential.

Lastly, another error source lies in the experimental identification
of the proper chiral $\sigma$-meson.  The precise location of the CEP
is rather sensitive to the mass value of the $\sigma$-meson
\cite{Schaefer:2008hk}, which we identified with the $f_0(500)$
resonance, whose mass is between $400-550\,\text{MeV}$
\cite{Agashe:2014kda}.  We have chosen $M_{\sigma} = 421$ MeV and for
larger values the CEP moves to larger chemical potential and smaller
temperature. Furthermore, it is still controversial whether the
$\sigma$-meson should be identified with the $f_0(500)$ or with the
$f_0(1370)$ resonance. It was argued e.g. in \cite{Parganlija:2010fz}
that the latter option might be favored and the $f_0(500)$ resonance
could be identified with a tetraquark state
\cite{Jaffe:1976ig}. However, within the FRG treatment it is not
possible to find a physically sensible set of initial conditions that
allow for an identification of the $\sigma$-meson with the
$f_0(1370)$, see also \cite{Eser:2015pka}.  If one would increase
$M_\sigma$ by increasing the initial value of
$\bar\lambda_{20,k=\Lambda}$ the UV masses of the other mesons would
decrease again. But as argued in Sec.~\ref{sec:ini}, these masses have
to be larger than the UV cutoff to ensure a suppression of the meson
fluctuation in the chirally symmetric regime.  We find that this
requirement excludes values larger than
$M_\sigma \gtrsim 600\,\text{MeV}$.  Hence, identifying the
$\sigma$-meson with the $f_0(1370)$ within our present setup is at
odds with initial conditions consistent with QCD.

\subsection{Pseudoscalar Mixing and the Axial Anomaly}\label{sec:mixing}

%%
%%%%%%%%%%%%%
\begin{figure}[t]
\begin{center}
  \includegraphics[width=.98\columnwidth]{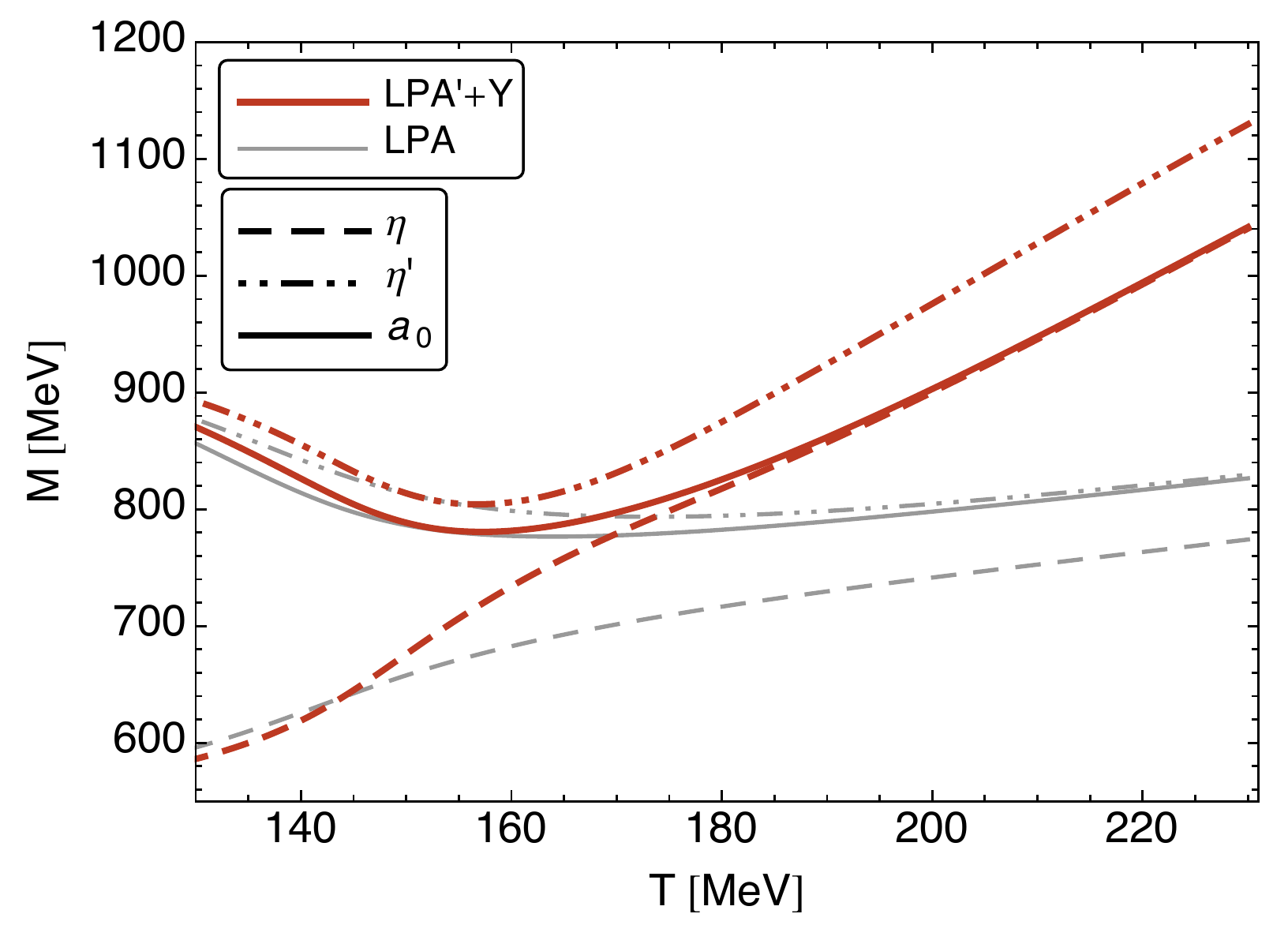}
  \caption{The masses of $\eta$-, $\eta^\prime$- and $a_0$-meson at
    large temperatures. While the $a_0$-meson (solid lines) degenerates
    with the $\eta^\prime$-meson in LPA (thin dotted-dashed gray
    line), it degenerates with the $\eta$-meson in LPA$^\prime$+Y (dashed
    red dark line).}\label{fig:mami}
\end{center}
\end{figure}
%%%%%%%%%%%%%
%%

\begin{figure}[t]
\begin{center}
  \includegraphics[width=.98\columnwidth]{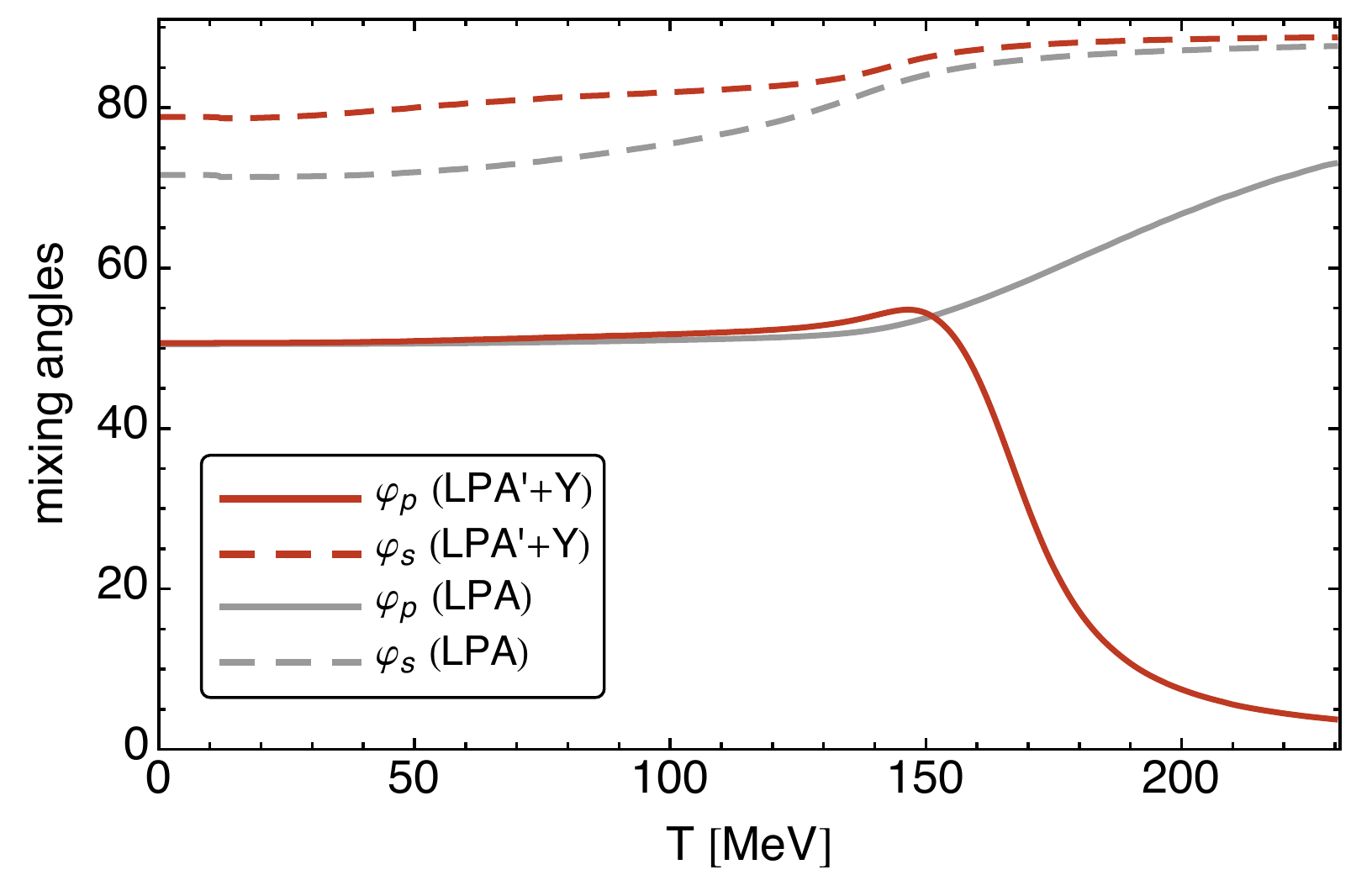}
  \caption{The scalar and pseudoscalar mixing angles $\varphi_{p,s}$
    as a function of temperature at vanishing density in
    LPA$^\prime$+Y (red lines) in comparison to the LPA results (gray
    lines).
    The $\eta$-meson becomes a purely light state at large
    temperatures and the $\eta^\prime$-meson a purely strange
    state.}\label{fig:ma}
\end{center}
\end{figure}

As already mentioned (cf.~Fig.~\ref{fig:masses}) the chiral partner of
the $a_0$-meson is the $\eta$-meson rather than the
$\eta^\prime$-meson in LPA$^\prime$+Y, which is displayed in more
detail in Fig.~\ref{fig:mami}. This finding is in contradistinction to
results found in a linear sigma model \cite{Lenaghan:2000ey} and in
quark-meson models \cite{Schaefer:2008hk, Tiwari:2013pg} in mean-field approximation,
as well as an FRG investigation with the quark-meson model in LPA
\cite{Mitter:2013fxa}. The reason is that in particular the
fluctuation-induced corrections to the meson propagators lead to the
opposite pseudoscalar mixing at large temperatures compared to
computations where these corrections are ignored. The pseudoscalar
mixing angle $\varphi_p$ describes the composition of the physical
$\eta$- and $\eta^\prime$-states in terms of the purely light $\eta_L$
and strange $\eta_S$ states, i.e., it describes the light and strange
quark content of the physical states, see \Eq{eq:plstrafo}.  The
mixing angles as a function of the temperature for vanishing density
are shown in Fig.~\ref{fig:ma} where the scalar (dashed lines) and the
pseudoscalar (solid lines) mixing angles obtained in LPA$^\prime$+Y
(red dark lines) are compared to the LPA results (gray light
lines). The scalar mixing angles $\varphi_s$ differ only mildly in
both truncations in contrast to the pseudoscalar mixing angle
$\varphi_p$ for temperatures above $T_c$. For a related DSE analysis
of the pseudoscalar mixing angle, see e.g. \cite{Horvatic:2007qs}.

If we ignore the running wave function renormalizations and the Yukawa
couplings, $\varphi_p$ grows for $T>T_c$ towards the ideal mixing
$\varphi_p \rightarrow 90^\circ$ (solid gray light line in
Fig.~\ref{fig:ma}). This implies that the $\eta$-meson becomes a
purely strange state and the $\eta^\prime$-meson a purely light state
when chiral symmetry is restored. Hence, the $\eta^\prime$ degenerates
with the light $a_0$-state at large temperatures in LPA. Similar for
the scalar sector: since $\varphi_s$ approaches ideal mixing in both
truncations, the $\sigma$-meson is a purely light and the $f_0$-meson
a purely strange state. Hence, the chiral partner of the $f_0$ is the
$\eta$-meson in LPA. However, since the strange condensate melts only
very slowly (cf.~Fig.~\ref{fig:mq}) we do not observe this
degeneration in the considered temperature range. This is in line with
results in the literature, e.g. \cite{Lenaghan:2000ey,
  Schaefer:2008hk} when quantum corrections to the classical
dispersion relation of the mesons, directly related to $Z_{\phi,k}$,
are neglected.

%%
%%%%%%%%%%%%%
%%%%%%%%%%%%%
%%

If we include these corrections by employing the LPA$^\prime$+Y
truncation, we find that $\varphi_p$ first slightly grows with
temperature until $T_c$ similar to the LPA result. $\varphi_p$ stays
close to 54$^\circ$ for $T \lesssim T_c$ meaning that the $\eta$ is an
almo
st ideal flavor octet state and $\eta^\prime$ an almost ideal
flavor singlet state. Above $T_c$ the pseudoscalar mixing angle
rapidly drops and finally approaches "anti-ideal" mixing
$\varphi_p \rightarrow 0^\circ$ at large temperatures (red dark solid
line in Fig.~\ref{fig:ma}). Hence, $\eta$ becomes a purely light state
with $a_0$ as its chiral partner and $\eta^\prime$ becomes purely
strange with $f_0$ as chiral partner. Again, we do not observe the
degeneration of $\eta^\prime$ with $f_0$ within the considered
temperature range due to the slow melting of the strange condensate
(cf.~Fig.~\ref{fig:mq}). In Tab.~\ref{tab:chiralpartners} we
summarized our findings. 

%%
%%%%%%%%%%%%%
\begin{table}[h]
\begin{center}
\begin{tabular}{ c c }
  \hline\hline mean-field/LPA/LPA+Y &\quad LPA$^\prime$ + Y \\
\hline
 $(\eta, f_0)$ &\quad $(\eta, a_0)$ \\
 $(\eta^\prime\!\!, a_0)$ &\quad $(\eta^\prime\!\!, f_0)$ \\
  \hline\hline
\end{tabular}
\end{center}
\caption{Chiral partners for different approximations.}
\label{tab:chiralpartners}
\end{table}
%%%%%%%%%%%%%
%%

%%
%%%%%%%%%%%%%
\begin{figure}[t]
\begin{center}
  \includegraphics[width=.98\columnwidth]{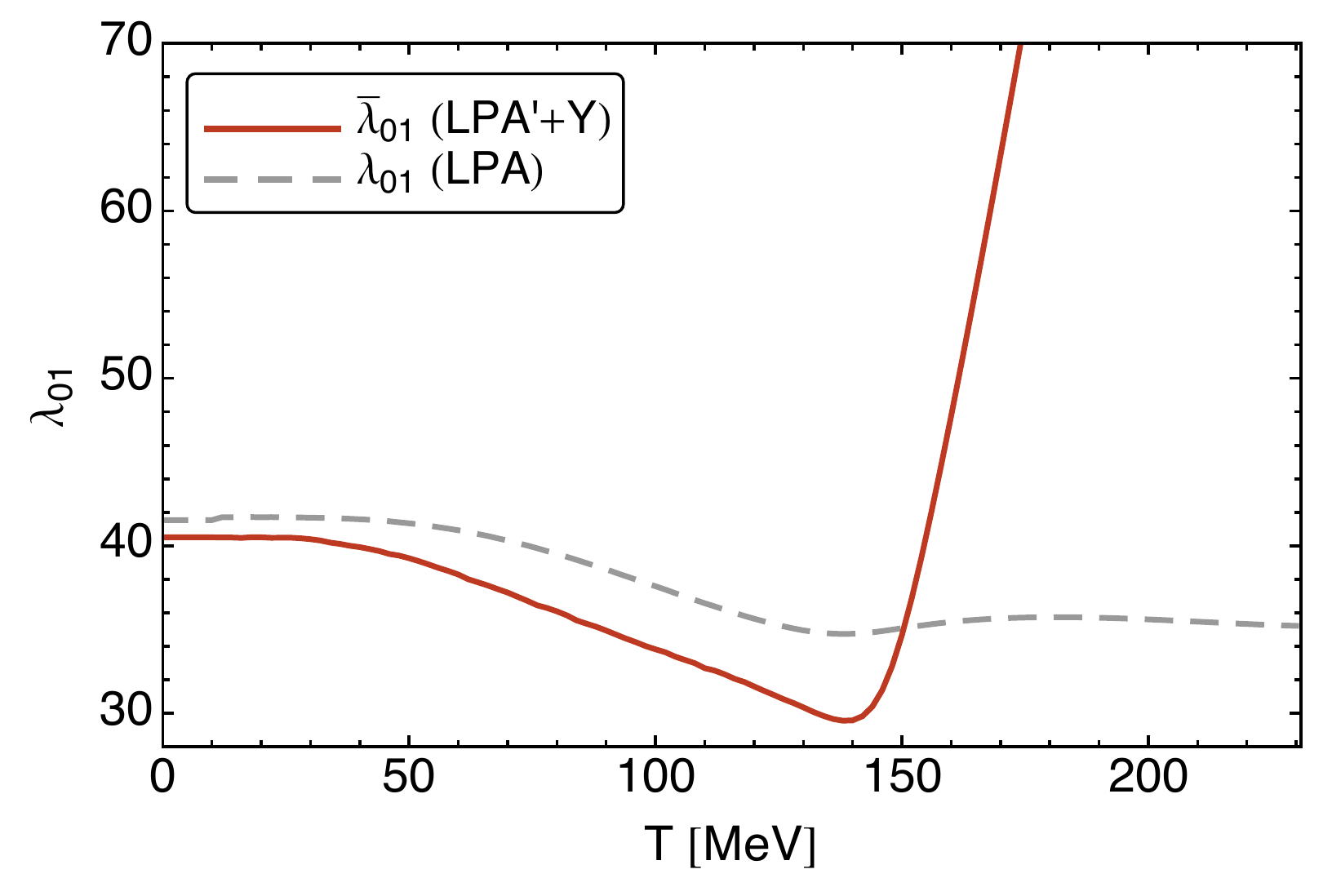}
  \caption{Temperature-dependence of the quartic meson coupling, which
    enters the pseudoscalar mixing angle, at vanishing density in
    LPA$^{\prime}$+Y (solid red line) and in LPA (dashed gray
    line).\label{fig:l01}}
\end{center}
\end{figure}
%%%%%%%%%%%%%
%%

We can further elucidate this phenomenon by examining the analytical
structure of the mixing angle. For convenience we restrict our
discussion to the 08-basis and call the pseudoscalar mixing angle
$\Theta_p$ that maps from the 08-basis to the physical one, see
App.~\ref{app:mixing} for details. Furthermore, we concentrate the
discussion without any limitations to vanishing density because the
mixing angle behaves similarly at finite $\mu$.  According to
\eq{eq:ls08} we have the relation
$\varphi_p = \Theta_p + \arctan \sqrt{2}$ which yields with
\Eq{eq:Theta} and the expressions in App.~\ref{app:mesmc} for $k=0$:
\begin{align}\label{eq:maexp}
\tan2 \Theta_p = 2 \sqrt{2}\, \frac{\big(\sqrt{2}
  \bar\sigma_{S}-\bar\sigma_{L}\big)\, \bar c -\big(2
  \bar\sigma_{S}^2-\bar\sigma_{L}^2\big) \bar\lambda_{01} }{\big(
  \sqrt{2} \bar\sigma_{S}+8 \bar\sigma_{L}\big)\, \bar c -\big(2
  \bar\sigma_{S}^2-\bar\sigma_{L}^2\big) \bar\lambda_{01}}\,.
\end{align}
The quartic meson coupling in the infrared related to
$\bar{\tilde\rho}_2$ is given by
$\bar\lambda_{01} = {\partial \bar U_{k=0}}/{\partial
  \bar{\tilde\rho}_2}\Big|_{\Phi_0}$.

Hence, $\Theta_p$ is solely driven by the condensates and the running
of $\bar\lambda_{01}$ because the KMT coupling $\bar c$ is a constant
as explained in Sec.~\ref{sec:pot}.  The temperature dependence of
$\bar\lambda_{01}$ in LPA$^\prime$+Y and LPA are shown in
Fig.~\ref{fig:l01}.  At vanishing temperature $\Theta_p$ is small and
negative since the light and strange condensates are almost identical,
$\bar{c}$ is constant and $\bar\lambda_{01,k} \approx 40$ in both LPA
and LPA$^\prime$+Y truncations cf. Fig.~\ref{fig:l01} and
Tab.~\ref{tab:init}. This means that
the $\eta^\prime$-state is mostly a $SU(3)$ singlet state at
$T\lesssim T_c$. With rising temperature, $\bar\sigma_{L}$ melts much
faster than $\bar\sigma_{S}$ and $\bar\lambda_{01}$ drops towards
$T_c$ which yields in an overall increase of the mixing angle. Thus,
for $T\lesssim T_c$ the mixing angle shows a similar behavior in both
truncations. But for $T>T_c$ the quartic coupling stays almost
constant in LPA while it grows rapidly in LPA$^\prime$+Y. The behavior
of the mixing angle in LPA is therefore dominated by the melting of
the condensates. In this case, the mixing angle tends to an ideal
mixing $\varphi_p \rightarrow 90^\circ$ since the light condensate
becomes negligible over the strange condensate and hence
\mbox{$\tan2 \Theta_p \rightarrow 2\sqrt{2}$} at large
temperature.

In LPA$^\prime$+Y, however, the quartic coupling $\bar\lambda_{01}$
grows much faster than the difference of the strange and light
condensate at $T>T_c$. As a consequence, \Eq{eq:maexp} is dominated by
$\bar\lambda_{01}$ and the mixing angle decreases with increasing
temperature in this regime. The denominator of \Eq{eq:maexp} decreases
until it vanishes around $T \approx 194\,\text{MeV}$ and
\mbox{$\tan2 \Theta_p \rightarrow 2\sqrt{2}$} at larger temperatures
because, again, the light condensate becomes negligible over the
strange condensate. Note, however, that by continuity the branch of
the complex tangent function has changed and thus the equation
\mbox{$\tan2 \Theta_p = 2\sqrt{2}$} implies an anti-ideal mixing
$\varphi_p = 0^\circ$. The rapid rise of $\bar\lambda_{01}$ in
Fig.~\ref{fig:l01} is directly linked to the rapid drop of
$Z_{\phi,k=0}$ in Fig.~\ref{fig:z}: Reparametrization invariance
implies the relation $\bar\lambda_{01} = \lambda_{01}/Z_{\phi,k=0}^2$.
Since these improvements are neglected in standard mean-field studies
of linear sigma and quark-meson models and in LPA with the FRG, such a
behavior of the mixing angle could not be observed in these cases.
In particular, in studies where quantum fluctuations are taken
  into account, it is crucial to capture the dynamical change of the
  relevant fluctuating degrees of freedom for a proper description of
  the phase transitions. 

  We have demonstrated that overestimating meson fluctuations in the
  vicinity of the chiral phase transition by not accounting for their
  proper decoupling from the physical spectrum above $T_c$ leads to a
  wrong assignment of the chiral partners of the ($\eta$,
  $\eta^\prime$) complex. The underlying mechanism is particularly
  lucid here. The behavior of $Z_{\phi,k=0}$ should be generic for QCD
  in the high-energy regime such that a similar behavior of the mixing
  angle is expected in full QCD.

  Interestingly, the same qualitative behavior of the pseudoscalar
  mixing angle has been observed in a Nambu--Jona-Lasinio (NJL) model
  study including the axial anomaly in mean-field approximation
  \cite{Costa:2005cz}. Moreover, similar findings were obtained for a
  Polyakov-loop augmented quark-meson model with vector mesons in an
  extended mean-field approximation where fermionic vacuum
  fluctuations are additionally considered
  \cite{Kovacs:2016juc}. These results are in general in line with our
  discussion above. However, we note that the main difference between
  \cite{Kovacs:2016juc} and \cite{Tiwari:2013pg} is the incorporation
  of vector mesons on the mean-field level in the former case. Yet,
  while $\varphi_p$ in \cite{Kovacs:2016juc} behaves qualitatively
  similar as in our LPA$^\prime$+Y truncation, it behaves as in LPA in
  \cite{Tiwari:2013pg}. This might be related
    to the different input parameters in the presence of vector
    mesons. In terms of fluctuations, however, recent studies beyond
    mean-field clearly indicate the irrelevance of the vector meson
    dynamics in Euclidean spacetime \cite{Rennecke:2015eba,
      Eser:2015pka, Jung:2016yxl}. Thus, even though different parameterizations of effective models can
    influence the value of the mixing angle, the
    dynamical effect identified in this work is expected to be
    relevant in any case.

    Finally, the observation
    of this phenomenon is directly related to the presence of the
    axial anomaly. In absence of the anomaly, i.e.~when $\bar c = 0$,
    the pseudoscalar mixing angle stays constant since
    $\tan2 \Theta_p = 2\sqrt{2}$ (cf.~\Eq{eq:maexp}). Thus, the
    non-trivial behavior of $\varphi_p$ as observed in
    Fig.~\ref{fig:ma} can only appear when $\bar{c} \neq 0$ i.e. in
    the presence of the axial anomaly. We remark that this observation
    may be an artifact of our model. A thorough study of the axial
    anomaly is beyond the scope of the present work.

%%%%%%%%%%%%%%%%%%%%%%%%%%%%%%%%%%%%%%%%%%%%%%%%%%%%%%%%%%%%%%
%%%%%%%%%%%%%%%%%%%%%%%%%%%%%%%%%%%%%%%%%%%%%%%%%%%%%%%%%%%%%%
\section{Conclusions and Outlook}\label{sec: concl}

We have investigated the influence of quantum and thermal fluctuations
on the chiral phase structure in the low-energy sector of QCD.
Motivated by the importance of mesonic degrees of freedom in QCD at
low energies, we employed an effective $N_f = 2+1$ quark-meson model
and incorporated quantum, thermal and density fluctuations by means of
the functional renormalization group.  The main focus was the
investigation of the impact of quark and meson fluctuation-induced
corrections to the classical quark and meson dispersion relations as
well as to quark-meson interactions. To this end, we allowed for
running light quark, strange quark and meson wave function
renormalizations $Z_{q/\phi,k}$ as well as running light and strange
quark Yukawa couplings $h_{q,k}$ and computed the resulting full
effective potential. Our two main findings were:

The considered fluctuations have an important and substantial impact
on the chiral phase boundary. The crossover transition is
significantly sharper as compared to computations that ignore the
running of the wave function renormalizations and the Yukawa couplings
(LPA truncation), see e.g. \cite{Schaefer:2006ds, Schaefer:2004en} for
similar two-flavor FRG investigations. Furthermore, the location of
the critical endpoint is pushed towards smaller density and larger
temperature. The overall quantitative effect on the chiral phase
boundary of these fluctuation-effects is substantial.

The pseudoscalar mixing angle, which determines the quark content of
  the pseudoscalar mesons, turned out to depend crucially on the
  elaborated interplay between quark and meson fluctuations. This is
intimately related to chiral symmetry restoration and the presence of
the axial anomaly. As a result, while the $\eta$- and
$\eta^\prime$-mesons are almost ideal flavor octet and singlet states
below $T_c$, the mixing angle approaches an anti-ideal flavor mixing
at temperatures above $T_c$ if all fluctuation corrections are taken
into account (LPA$^{\prime}$+Y truncation). The $\eta$-meson
  becomes a purely light quark composite and $\eta^\prime$ purely
  strange. In absence of the axial anomaly, the $\eta^\prime$
  degenerates with the pions above $T_c$ and therefore becomes a light
  quark state. As a consequence, a rearrangement of the chiral meson
partner $(a_0,\eta)$ and $(f_0,\eta^\prime)$ occurs in the presence of
the axial anomaly. However, we consolidated this observation by
clarifying its direct connection to the running of the meson wave
function renormalization $Z_{\phi,k}$.  We thus expect that our
findings are also qualitatively incorporated in full QCD and describe
more realistically the relevant physical superposition of gluon and
hadron effects including their mutual back-reaction in particular in
the transition region.

Of course there is still space for further improvements: the neglected
gluon fluctuations can at least effectively included in this framework
by a non-vanishing temporal gluon background yielding Polyakov-loop
augmented quark-meson model truncations \cite{Schaefer:2007pw,
  Schaefer:2009ui, Karsch:2010hm}. This enables additionally the
analysis of deconfinement issues in the same setup as discussed in
Sec.~\ref{sec:pb}. Another important aspect is the unknown
significance of baryon degrees of freedom at finite densities and
their influence on the location of a possible critical endpoint. A
first exploratory study in a DSE framework wherein quark and gluon
fluctuations are taken into account indicates that baryons seems to
have only little impact on the location of the CEP in the phase
diagram \cite{Eichmann:2015kfa}.

Aiming at a more accurate description of QCD along the lines of
e.g. \cite{Herbst:2010rf, Mitter:2014wpa, Braun:2014ata} as well as
phenomenological applications such as baryon and strangeness
fluctuations are obvious directions for future work.

%%%%%%%%%%%%%%%%%%%%%%%%%%%%%%%%%%%%%%%%%%%%%%%%%%%%%%%%%%%%%%%

\vspace{0.1cm} {\it Acknowledgments -} We thank Mario Mitter, Jan
M. Pawlowski and Simon Resch for discussions and collaboration on
related projects. This work has been supported by the FWF grant P24780-N27,
the Helmholtz International Center for FAIR within the LOEWE
program of the State of Hesse and by the DFG Collaborative Research Centre "SFB 1225 (ISOQUANT)".

%%%%%%%%%%%%%%%%%%%%%%%%%%%%%%%%%%%%%%%%%%%%%%%%%%%%%%%%%%%%%%%%
\begin{appendix}

%%%%%%%%%%%%%%%%%%%%%%%%%%%
\section{Two-Dimensional Fixed Background Taylor Expansion}\label{app:taylor}

There are various ways to solve the flow equation for the field
dependent effective potential $U_k(\rho_1,\tilde \rho_2)$. Most
common choices are either to discretize the potential in field space
or perform a Taylor expansion. While the former provides very accurate
information about the global potential structure, the latter only
gives accurate local information about the potential and its
derivatives in the vicinity of the expansion point. Even though the
full global information is only hardly accessible within a Taylor
expansion, it is numerically less extensive.  While an expansion about
the running minimum of the effective potential seems to be the most
natural choice, its convergence properties are rather poor as has been
demonstrated in \cite{Pawlowski:2014zaa}.  A static expansion about a
scale-independent point which coincides with the IR-minimum of the
effective potential is numerically more stable and converges rapidly.

Here, we generalize this static background Taylor expansion to two
dimensions. We expand the effective potential
$\tilde{U}_k(\rho_1,\tilde\rho_2)$ about two scale-independent
expansion points $\kappa_1$, $\kappa_2$ to a given order $N$
\begin{align}\label{eq:potexp}
U_k(\rho_1,\tilde\rho_2) =
  \sum_{i,j=0}^N\frac{\lambda_{ij,k}}{i!j!}(\rho_1-\kappa_1)^i(\tilde\rho_2-\kappa_2)^j
\end{align}
with $\lambda_{00,k}=0$.  There are various way to define this
expansion: while an expansion directly in powers of the chiral
invariants seems to be the easiest choice, an expansion in powers of
the fields is more natural. We also found the latter to be numerically
more stable. In the present case $N$ fixes the order of the fields,
i.e., $i,j \in \{0,1,\dots,N+1\}$ with $0 < i\!+\!2 j \leq N\!+\!1$.
This corresponds to an expansion up to $\phi^{2(N + 1)}$
\footnote{Note that $\rho_1 \sim \phi^2$ and
  $\tilde\rho_2 \sim \phi^4$.}.

The RG-invariant expansion points
$\bar \kappa_{1,k} = Z_{\phi,k} \kappa_1$ and
$\bar \kappa_{2,k} = Z_{\phi,k}^2 \kappa_2$ then lead to the following
running of these points:
\begin{align}
\partial_t\bar \kappa_{1,k} = -\eta_{\phi,k} \bar \kappa_{1,k}\,,\qquad \partial_t\bar \kappa_{2,k} = -2\eta_{\phi,k} \bar \kappa_{2,k}\,.
\end{align}
The initial values for these flow equations are chosen such that their
IR values coincide with the minimum of the effective potential in
light and strange dimension more precisely with the chiral invariants
at the physical point
\begin{align}\label{eq:rhosigmarel}
\begin{split}
\bar\rho_{1,k} &= \frac{1}{2}\left(\bar\sigma_{L,k}^2 +\bar\sigma_{S,k}^2\right)\,,\\
 \tilde{\bar\rho}_{2,k} &= \frac{1}{24}\left(\bar\sigma_{L,k}^2 -2 \bar\sigma_{S,k}^2\right)^2\,,
\end{split}
\end{align}
where we demand
\begin{align}
\bar \kappa_{1,0} = \bar\rho_{1,0}\,,\quad\text{and}\quad  \bar \kappa_{2,0} = \tilde{\bar\rho}_{2,0}\,.
\end{align}
To extract the condensates $\bar{\sigma}_{L,k}$ and
$\bar{\sigma}_{S,k}$ one has to invert \Eq{eq:rhosigmarel}.
Even though this is not unique in general, at the physical point we
always have $\bar\sigma_{L,k}\leq\sqrt{2}\bar\sigma_{S,k}$ with the
unique solution
\begin{align}\label{eq:condinv}
\begin{split}
\bar\sigma_{L,k} &= \sqrt{\frac{2}{3}\left( 2 \bar\rho_{1,k}  -\sqrt{6
      \tilde{\bar\rho}_{2,k} } \right)}\,,\\ 
\bar\sigma_{S,k} &= \sqrt{\frac{2}{3}\left(\bar\rho_{1,k}  + \sqrt{6
      \tilde{\bar\rho}_{2,k} } \right)}\,. 
\end{split}
\end{align}
In practice, we fine-tune the initial conditions for the expansion
points $\bar \kappa_{1,k=\Lambda}$ and $\bar \kappa_{2,k=\Lambda}$ such
that
\begin{align}
\begin{split}
\bar\sigma_{L,k=0} \,<\, \sqrt{\frac{2}{3}\left( 2 \bar\kappa_{1,k=0}
    -\sqrt{6 \kappa_{2,k=0} } \right)} \,&\leq\, (1+\epsilon)\,
\bar\sigma_{L,k=0} \,,\\ 
\bar\sigma_{S,k=0} \,<\,\sqrt{\frac{2}{3}\left(\bar\kappa_{1,k=0}
    +\sqrt{6 \kappa_{2,k=0} } \right)} \,&\leq\,
(1+\epsilon)\,\bar\sigma_{S,k=0}\,, 
\end{split}
\end{align}
where we typically choose $\epsilon$ of the order of $5\cdot 10^{-4}$.
For the numerical stability of the expansion it is crucial that the
expansion points do not lie in the convex region of the effective
potential for $k \rightarrow 0$. However, being close to the physical
minimum is indispensable if the truncation of the effective action
contains field-independent parameters since the parameters obtain a
physical meaning only when they are defined at the physical point in
the IR.  Here, this is the case for the wave function renormalizations
and the Yukawa couplings. If one would also expand these terms, one
only needs to choose the expansion points such that the physical point
lies within the radius of convergence of the expansion, which would
allow for much larger values for $\epsilon$. For more details see
\cite{Pawlowski:2014zaa}.

The flow equations of the expansion coefficients in \Eq{eq:potexp}
are:
\begin{align}
\partial_t \bar\lambda_{nm,k}= (n+2m)\eta_{\phi,k}\bar\lambda_{nm,k} +
  \frac{\partial^{n+m} \partial_t 
  U_k(\rho_1,\tilde\rho_2)}{\partial\rho_1^n\partial\tilde\rho_2^m}\bigg|_{\kappa_1,\kappa_2}\,,  
\end{align}
where $\partial_t U_k(\rho_1,\tilde\rho_2)$ is given by
\Eq{eq:uflow}. We have used $N=4$ (corresponding to a
$\phi^{10}$-expansion) throughout this work.  We found that our
results are independent of $N$ for $N\geq 3$ and
$\mu \lesssim 250\,\text{MeV}$. At larger $\mu$, the distance between
the local minima of the effective potential in the region of the
first-order phase transition appears to be larger than the radius of
convergence of the expansion at $N=4$. Hence, we do not resolve the
phase boundary at larger chemical potential.

We want to end with a discussion of the field dependence of the flow
equation. In general, it can be a complicated non-linear function of
various invariants, including potential non-analytical terms. This can
be inferred, e.g., from \Eq{eq:condinv} and was also discussed in
\cite{Patkos:2012ex}. There, it was shown that higher powers of an
expansion of the effective potential in terms of the invariants
$\rho_1$ and $\tilde\rho_2$ indeed contain fractional powers of the
invariants. It was argued that they arise due to the omission of the
third invariant. However, the fast convergence of our expansion
suggests that we do not suffer from any non-analyticities within the range
of parameters we consider here. Furthermore, we are able to accurately
reproduce the results of \cite{Mitter:2013fxa}, wherein the flow of the
$U(3)\!\times\! U(3)$-symmetric part of the effective potential was
solved on a two-dimensional grid of the two invariants $\rho_1$ and
$\tilde\rho_2$ in the LPA truncation. Aside from the physical
requirement that the effective potential is a function of the chiral
invariants, no assumptions are made about the functional form of the
potential. This method captures the full field dependence of the
effective potential, including potential non-analyticities. Owing to
the fast convergence of our expansion and the reproduction of the
results in \cite{Mitter:2013fxa} we are confident that we accurately
solve the flow of the effective potential at the physical point within
our truncation.

Furthermore, we again want to emphasize that the
  assumption that the anomaly coefficient $\bar c$ is not running is
part of our truncation. In principle, one could extract a flow of this
coupling from the effective potential by considering an appropriate
three-point function. This is beyond the scope of the present work. An
elegant way to disentangle and extract the flow of the anomaly
coefficient as well as to close the set of flow equations in the LPA
truncation of a linear sigma model is presented in
\cite{Fejos:2016hbp}.

%%%%%%%%%%%%%%%%%%%%%%%%%%%
\section{Threshold Functions}\label{app:rucou}

In this appendix the regulators and threshold functions of the flow
equations are specified.  We restrict ourselves to $d=4$
dimensions. For the light and strange quarks as well as the mesons the
following three-dimensional regulators are used so that only the
spatial momenta are regulated
\begin{align}\label{eq:regs}
\begin{split}
R_k^q(\vec{p}) &= Z_{q,k}\, \vec{\gamma}\vec{p}\, r_F\left(\vec{p}^2/k^2\right)\,\\
R_k^\phi(\vec{p}^2) &= Z_{\phi,k}\, \vec{p}^2\, r_B\left(\vec{p}^2/k^2\right)
\end{split}
\end{align}
with the optimized fermionic $r_F(x)$ and bosonic $r_B(x)$ regulator shape functions
\cite{Litim:2006ag, Litim:2001up, Litim:2000ci} ($x=\vec{p}^2/k^2$)
\begin{align}
\begin{split}
r_F(x) &= \left(\frac{1}{\sqrt{x}}-1\right) \Theta(1-x)\,,\\
r_B(x) &= \left(\frac{1}{x}-1\right) \Theta(1-x)\,.
\end{split}
\end{align}
As mentioned in Sec.~\ref{sec:frg}, the regulator ensures both UV and
IR regularity of our theory. Finite temperature is implemented via the
imaginary time formalism. Hence, the loop frequency integration is a
sum over Matsubara frequencies $\omega_n = 2\pi n T$ and
$\nu_n = 2 \pi (n+1/2) T$ for bosons and fermions respectively. At
large temperatures the meson contributions reduce to the
three-dimensional case due to dimensional reduction. Since the quark
propagators vanish for $p_0\rightarrow\infty$, the quark contributions
are also UV regular.

With these regulators, the scalar parts without the tensor structure
of the quark and meson propagators read
\begin{align}
\begin{split}
\bar G_{q,k}(\nu_n,\vec{p}) &=
\frac{\Theta(k^2-\vec{p}^2)}{Z_{q,k}\big((\nu_n+i\mu)^2+k^2\big)+m_{q,k}^2}\\ 
&\quad+
\frac{\Theta(\vec{p}^2-k^2)}{Z_{q,k}\big((\nu_n+i\mu)^2+\vec{p}^2\big)+m_{q,k}^2}\,,\\ 
\bar G_{\phi,k}(\omega_n,\vec{p}) &= 
\frac{\Theta(k^2-\vec{p}^2)}{Z_{\phi,k}\big(\omega_n^2+k^2\big)+m_{\phi,k}^2}\\ 
&\quad+ 
\frac{\Theta(\vec{p}^2-k^2)}{Z_{\phi,k}\big(\omega_n^2+\vec{p}^2\big)+m_{\phi,k}^2}\,. 
\end{split}
\end{align}
In the spirit of the low-momentum expansion all flow equations are
evaluated at vanishing spatial external momenta,
cf.~e.g. \Eq{eq:pext}. As a consequence, only the parts of the
propagators proportional to $\Theta(k^2-\vec{p}^2)$ will contribute to
the final flow equations. Hence, in the following we will use the
dimensionless RG-invariant propagators
\begin{align}
\begin{split}
\tilde G_{q,k}(p_0) &= \frac{1}{(p_0+i\mu)^2/k^2 + 1 + \bar m_{q,k}^2}\,,\\
\tilde G_{\phi,k}(p_0) &= \frac{1}{p_0^2/k^2 + 1 + \bar m_{\phi,k}^2}\,,
\end{split}
\end{align}
which directly enter the loop frequency summations.

The functions $l_0^{(B/F)}$ in \Eq{eq:uflow} are related to
the bosonic/fermionic loops and are defined as 
\begin{align}\label{eq:l0b}
&l_0^{(B)}(\bar{m}_{\phi,k}^2,\eta_{\phi,k})\\ \nonumber
&=\frac{T}{2 k}\sum_{n\in \mathbb{Z}} \int\!dx x^{\frac{3}{2}}\left( \partial_t r_B(x)-\eta_{\phi,k} r_B(x)\right) \tilde G_{\phi,k}(\omega_n)\\ \nonumber
&=\frac{2}{3}\frac{k}{E_{\phi,k}}
\left(1-\frac{\eta_{\phi,k}}{5}\right) \left(\frac{1}{2}+n_B(E_{\phi,k})\right)\,,
\end{align}
for mesons and
\begin{align}\label{eq:l0f}
&l_0^{(F)}(\bar{m}_{q,k}^2,\eta_{q,k})\\ \nonumber
&=\frac{T}{k}\sum_{n\in \mathbb{Z}} \int\!\!\!dx x^{\frac{3}{2}}\left(\partial_t r_F(x)-\eta_{q,k} r_F(x)\right) (1+r_F(x))\tilde G_{q,k}(\nu_n)\\ \nonumber
&=\frac{1}{3}\frac{k}{E_{q,k}}\left(1-\frac{\eta_{q,k}}{4}\right) \Big[1-n_F(E_{q,k}-\mu)-
n_F(E_{q,k}+\mu)\Big]\,,
\end{align}
for quarks with the usual Bose and Fermi distributions
\begin{align}
\begin{split}
n_B(E) = \frac{1}{e^{E/T}-1}\,,\quad n_F(E) = \frac{1}{e^{E/T}+1}\,,
\end{split}
\end{align}
and the scale dependent quasiparticle energies
\begin{align}
E_{\Phi,k} = k \sqrt{1+\bar m_{\Phi,k}^2}\, \qquad\text{with} \qquad\bar
  m_{\Phi,k} = M_{\Phi,k}/k\,.
\end{align}
The threshold functions $L_{(1,1)}$ that appear in the flows of the
Yukawa couplings, \eq{eq:hlflow} and \eq{eq:hsflow}, are related to
the triangle diagrams as shown in Fig.~\ref{fig:dcont}(a). They are defined by
\begin{align}\label{eq:l11}
\begin{split}
&L_{(1,1)}\left(\bar{m}_{q,k}^2,\bar{m}_{\phi,k}^2,\eta_{q,k},\eta_{\phi,k};p_0\right)\\
&=\frac{T}{2k}\sum_{n\in\mathbb{Z}}\int\! dx x^{\frac{3}{2}}
\bigg[\left(\partial_t r_B(x)-\eta_{\phi,k} r_B(x)\right)\\
&\quad\times \tilde G_{q,k}(\nu_n+p_0)  \tilde G_{\phi,k}^2(\omega_n)+2\big(1+r_F(x)\big)\\
&\quad\times\left(\partial_t r_F(x)-\eta_{q,k} r_F(x)\right)
 \tilde G_{q,k}^2(\nu_n+ p_0)  \tilde G_{\phi,k}(\omega_n)  \bigg]\,.
\end{split}
\end{align}
As already mentioned we evaluate all flows at vanishing external
spatial momenta and only the external frequency $p_0$ is
retained. Owing to the optimized 3d regulators, the spatial
integration can be performed analytically
\begin{align}
\begin{split}
&L_{(1,1)}\left(\bar{m}_{q,k}^2,\bar{m}_{\phi,k}^2,\eta_{q,k},\eta_{\phi,k};p_0\right)\\ 
&=\frac{2}{3} \left[\!
  \left(1-\frac{\eta_{\phi,k}}{5}\right)\!\textit{FB}_{(1,2)}^{(q,\phi)}+\!\left(1-\frac{\eta_{q,k}}{4}\right)\!\textit{FB}_{(2,1)}^{(q,\phi)}
\right]\,, 
\end{split}
\end{align}
where we have suppressed  the arguments of 
$\textit{FB}_{(1,2)}^{(q,\phi)} = \textit{FB}_{(1,2)}(\bar
m_{q,k}^2,\bar m_{\phi,k}^2;p_0)$.
The remaining Matsubara summation can also be performed
analytically with the result
\begin{align}\label{eq:fb11}
&\textit{FB}_{(1,1)}(\bar m_{q,k}^2,\bar m_{\phi,k}^2;p_0)\\ \nonumber
&=  \frac{T}{k}\operatorname{Re}\left[\sum_{n\in \mathbb{Z}}\tilde
  G_{q,k}(\nu_n+p_0)\tilde G_{\phi,k}(\omega_n)\right]\\ \nonumber 
&= \text{Re}\,\Bigg[\frac{k}{2 E_{\phi,k}} n_B(E_{\phi,k})
  \frac{k^2}{E_{q,k}^2-\big( E_{\phi,k}- i p_0 + \mu\big)^2}\\
  \nonumber 
&\quad +  \frac{k}{2 E_{\phi,k}} \big(n_B(E_{\phi,k})+1\big)
  \frac{k^2}{E_{q,k}^2-\big(E_{\phi,k}+ i p_0 - \mu \big)^2}\\
  \nonumber 
&\quad - \frac{k}{2 E_{q,k}} n_F(E_{q,k}-\mu)
  \frac{k^2}{E_{\phi,k}^2-\big( E_{q,k}+ i p_0 - \mu \big)^2}\\
  \nonumber 
&\quad - \frac{k}{2 E_{q,k}} \big(n_F(E_{q,k}+\mu)-1\big)
  \frac{k^2}{E_{\phi,k}^2-\big( E_{q,k}- i p_0 + \mu
  \big)^2}\Bigg]\,. 
\end{align}
The functions $\textit{FB}_{(n,m)}^{(q,\phi)}$ correspond to the
Matsubara sum of $n$ quarks propagators and $m$ meson propagators and
can easily be obtained from the function
$\textit{FB}_{(1,1)}^{(q,\phi)}$ via derivatives
\begin{align}\label{eq:fbnm}
\textit{FB}_{(n,m)}^{(q,\phi)} = \frac{(-1)^{n+m-2}}{(n-1)! (m-1)!}
  \frac{\partial^{n+m-2} \textit{FB}_{(1,1)}^{(q,\phi)}}{\partial
  (\bar m_{q,k}^2)^{n-1}  \partial (\bar m_{\phi,k}^2)^{m-1}}\,, 
\end{align}
with $n,m \geq 1$. Note that these functions also enter the light and
strange quark anomalous dimensions Eqs.~\eq{eq:etal} and \eq{eq:etas}.

Due to the complex frequency arguments, the real part in \Eq{eq:fb11}
has to be taken in order to ensure real-valued Yukawa couplings and
quark anomalous dimensions which also avoids an unphysical
complex-valued effective potential.  For fermionic frequencies the
choice $p_{0,\text{ext}} = 0$ is excluded.  However, as discussed in
the main text, we choose $p_{0,\text{ext}} = \pi T - i \mu$, see
\Eq{eq:pext}.  The inclusion of the fully frequency-dependent quark
anomalous dimension into the effective potential results in a
manifestly real effective potential. But this is beyond the scope of
the present work and we refer to \cite{Fu:2016tey} for further
details.

The functions $\textit{BB}_{(n,m)}^{(\phi_1,\phi_2)}$ in the meson
anomalous dimension \eq{eq:etaphi} encodes the Matsubara summation of
loops with two different meson propagators and are defined as:
\begin{align}
\begin{split}
&\textit{BB}_{(1,1)}(\bar{m}_{\phi_1,k}^2,\bar{m}_{\phi_2,k}^2) \,=\, \frac{T}{k}\sum_{n\in \mathbb{Z}} \tilde G_{\phi_1}(\omega_n) \tilde G_{\phi_2}(\omega_n)\\
&= \frac{k^3}{(E_{\phi_2,k}^2-E_{\phi_1,k}^2)\, E_{\phi_1,k}}\left(n_B(E_{\phi_1,k})+\frac{1}{2}\right)\\
&\quad+\frac{k^3}{(E_{\phi_1,k}^2-E_{\phi_2,k}^2)\,E_{\phi_2,k}}\left(n_B(E_{\phi_2,k})+\frac{1}{2}\right)\,.
\end{split}
\end{align}
This definition is only valid for
$\bar{m}_{\phi_1,k}\neq\bar{m}_{\phi_2,k}$. In the case of a mass
degeneracy
$\bar{m}_{\phi_1,k} = \bar{m}_{\phi_2,k} \equiv \bar{m}_{\phi,k}$ one
has to use
\begin{align}
\begin{split}
&\textit{BB}_{(1,1)}(\bar{m}_{\phi,k}^2,\bar{m}_{\phi,k}^2) = -\frac{\partial}{\partial \bar{m}_{\phi,k}^2}\left[\frac{k}{E_{\phi,k}} \left(n_B(E_{\phi_2,k})+\frac{1}{2}\right)\right]\,.
\end{split}
\end{align}
Accordingly, for $n$ propagators of $\phi_1$ and $m$ propagators of $\phi_2$ we find
\begin{align}\label{eq:fbnm}
\textit{BB}_{(n,m)}^{(\phi_1,\phi_2)} = \frac{(-1)^{n+m-2}}{(n-1)!
  (m-1)!} \frac{\partial^{n+m-2}
  \textit{BB}_{(1,1)}^{(\phi_1,\phi_2)}}{\partial (\bar
  m_{\phi_1,k}^2)^{n-1}  \partial (\bar m_{\phi_2,k}^2)^{m-1}}\,. 
\end{align}
The Matsubara summation of loops with several identical fermions also
enters in \Eq{eq:etaphi} and is encoded in:
\begin{align}
\begin{split}
&\textit{F}_{(1)}(\bar{m}_{q,k}^2) \,=\, \frac{T}{k}\sum_{n \in \mathbb{Z}} \tilde G_q(\nu_n)\\
& = \frac{k}{2 E_{q,k}}\Big[1-n_F(E_{q,k}-\mu)-n_F(E_{q,k}+\mu)\Big]\,,
\end{split}
\end{align}
and
\begin{align}
\textit{F}_{(n)}^{\,(q)}&=\frac{(-1)^{n-1}}{(n-1)!}\frac{\partial^{n-1} \textit{F}_{(1)}^{\,(q)}}{\partial (\bar m_{q,k}^2)^{n-1}}\,.
\end{align}
Note that this function is implicitly contained in the threshold
function $l_{n}^{(F)}$ that appears in the flow of the effective
potential.

%%%%%%%%%%%%%%%%%%%%%%%%%%%
\section{Mixing Angles}\label{app:mixing}

As discussed in Sec.~\ref{sec:model}, the mixing angles $\varphi_s$
and $\varphi_p$ relate the purely light and strange scalar and
pseudoscalar mesons to the physical ones, which are defined as mass
eigenstates.  Formally, the mixing angles $\varphi_{s/p}$ rotate
mesons from the LS- to the physical basis.

On the other hand, the scalar and pseudoscalar meson nonets can also
be defined directly from the $U(N_f)$ generators
$T_{0},\dots,T_{N_f^2-1}$, see \Eq{eq:Sigma} which we call the
08-basis. The rotation from the 08- to the physical basis is
accomplished by the mixing angles $\Theta_{s/p}$ via
\begin{align}\label{eq:p08trafo}
\begin{split}
\begin{pmatrix}f_0 \\ \sigma \end{pmatrix} &= \begin{pmatrix} \cos\Theta_s & - \sin\Theta_s \\ \sin\Theta_s & \cos\Theta_s \end{pmatrix} \begin{pmatrix}\sigma_8 \\ \sigma_0 \end{pmatrix}\,,\\
\begin{pmatrix}\eta \\ \eta^\prime \end{pmatrix} &= \begin{pmatrix} \cos\Theta_p & - \sin\Theta_p \\ \sin\Theta_p & \cos\Theta_p \end{pmatrix} \begin{pmatrix}\eta_8 \\ \eta_0 \end{pmatrix}\,.
\end{split}
\end{align}
In summary, for the angle $\varphi$ the Hessian has to be computed in
the LS-basis and for $\Theta$ it has to be computed in the
08-basis. In general, the Hessian in the basis $b$, $\bar H^{(b)}$, is
block-diagonal,
\begin{align}\label{eq:hessian}
\bar H^{(b)} = \begin{pmatrix} \bar H^{(b)}_s & 0 \\ 0 & \bar H^{(b)}_p  \end{pmatrix}\,,
\end{align}
where $\bar H^{(b)}_{s/p}$ denote the scalar/pseudoscalar blocks. They are defined as
\begin{align}\label{eq:Hblock}
  \bar H^{(b)}_{s/p,ij} = \frac{\partial^2}{\partial\bar
  \phi_{s/p,i}^{(b)} \partial\bar \phi_{s/p,j}^{(b)} } \bar{\tilde
  U}_k\left(\bar\Sigma^{(b)},(\bar\Sigma^{(b)})^\dagger\right)\,. 
\end{align}
The entries of the Hessian in the 08-basis, $\bar H^{(08)}$ are explicitly given in App.~\ref{app:mesmc}.

With these definitions, we find for the scalar and pseudoscalar mixing
angles in the 08-basis 
\begin{align}\label{eq:Theta}
\Theta_{s/p} = \frac{1}{2} \arctan\left(\frac{2 \bar H_{s/p,19}^{(08)}}{ \bar H_{s/p,11}^{(08)}- \bar H_{s/p,99}^{(08)}}\right)\,,
\end{align}
where we have used $\bar \phi^{(08)}_{s,1/9} = \sigma_{0/8}$ and
$\bar\phi^{(08)}_{p,1/9} = \eta_{0/8}$. 
Similarly, in the LS-basis we find
\begin{align}\label{eq:varphi}
\varphi_{s/p} = \frac{1}{2} \arctan\left(\frac{2 \bar H_{s/p,19}^{(\text{LS})}}{ \bar H_{s/p,99}^{(\text{LS})}- \bar H_{s/p,11}^{(\text{LS})}}\right)\,,
\end{align}
with $\bar\phi^{(\text{LS})}_{s,1/9} = \sigma_{L/S}$ and $\bar\phi^{(\text{LS})}_{p,1/9} = \eta_{L/S}$. 
Finally, by means of \Eq{eq:ls08} a direct relation between the angles $\varphi$ and $\Theta$
can be found as follows
\begin{align}
\begin{split}
\sin\varphi_{s/p} &= \sqrt{\frac{1}{3}} \sin\Theta_{s/p} + \sqrt{\frac{2}{3}}\cos\Theta_{s/p}\,,\\
\cos\varphi_{s/p} &= \sqrt{\frac{1}{3}} \cos\Theta_{s/p} - \sqrt{\frac{2}{3}}\sin\Theta_{s/p}\,.
\end{split}
\end{align}
Note that some caution is advised with the definition of the mixing
angles in Eqs.~\eq{eq:Theta} and \eq{eq:varphi}. The denominator can
become zero and, by continuity of the mixing angle, the branch of the
arctan changes. This is relevant for the pseudoscalar mixing angle at
large temperatures.

%%%%%%%%%%%%%%%%%%%%%%%%%%%

\section{Meson Masses and Couplings}\label{app:mesmc}

The mesons masses are given by the eigenvalues of the Hessian $\bar H$
of the effective potential defined in Eqs.~\eq{eq:hessian} and
\eq{eq:Hblock}. If we choose the physical basis for the meson fields,
the Hessian is diagonal by definition. In any other case,
diagonalization is necessary. In the following, we work in the
singlet-octet basis (08-basis), which implies that the scalar and
pseudoscalar meson nonets are
\begin{align}
\begin{split}
\bar\phi^{(08)}_{s} =
(\bar\sigma_0,\bar\sigma_1,\dots,\bar\sigma_7,\bar\sigma_8 )\,,\\
 \bar\phi^{(08)}_{p} = (\bar\eta_0,\bar\pi_1,\dots,\bar\pi_7, \bar\eta_8)\,.
\end{split}
\end{align}
The squared meson masses $M_{\phi,k}^2$ can then be expressed in terms
of the entries of $\bar H^{(08)}_s$ and $\bar H^{(08)}_p$ defined in
\Eq{eq:Hblock} as well as the scalar and pseudoscalar mixing angles
$\Theta_s$ and $\Theta_p$ discussed in App.~\ref{app:mixing}. Since
$\bar\sigma_0$ and $\bar\sigma_8$ are directly linked to the
condensates $\bar\sigma_{L,k}$ and $\bar\sigma_{S,k}$ via a rotation,
\Eq{eq:ls08}, we express the $\bar H^{(08)}_{s/p,ij}$ solely in terms
of the condensates as well as derivatives of the effective potential
with the abbreviation
\begin{align}
\frac{\partial^{n+m} \bar U(\bar \rho_1,\bar{\tilde \rho}_2)}{\partial \bar \rho_1^n \partial\bar{\tilde \rho}_2^m} \equiv \bar U^{(n,m)}\,.
\end{align}
We first summarize the relations between the squared physical masses
and the entries of the Hessian. For the scalar meson nonet they read:
\begin{widetext}
\begin{align}
M_{f_0,k}^2 &= \cos^2\Theta_s\, \bar H_{s,99}^{(08)} + \sin^2\Theta_s\, \bar H_{s,11}^{(08)} - 2 \cos\Theta_s \sin\Theta_s\, \bar H_{s,19}^{(08)}\,,\\
M_{\sigma,k}^2 & = \sin^2\Theta_s\, \bar H_{s,99}^{(08)} + \cos^2\Theta_s\, \bar H_{s,11}^{(08)} + 2 \cos\Theta_s \sin\Theta_s\, \bar H_{s,19}^{(08)}\,,\\
M_{a_0,k}^2 &= \bar H_{s,ii}^{(08)}\quad i=2,3,4\,,\\
M^2_{\kappa,k} &= \bar H_{s,ii}^{(08)}\quad i=5,6,7,8\,.
\end{align}
and equivalently for the pseudoscalar nonet:
\begin{align}
M_{\eta,k}^2 &= \cos^2\Theta_p\, \bar H_{p,99}^{(08)} + \sin^2\Theta_p\, \bar H_{p,11}^{(08)} - 2 \cos\Theta_p \sin\Theta_p\, \bar H_{p,19}^{(08)}\,,\label{eq: meta}\\
M_{\eta^\prime,k}^2 & = \sin^2\Theta_p\, \bar H_{p,99}^{(08)} + \cos^2\Theta_p\, \bar H_{p,11}^{(08)} + 2 \cos\Theta_p \sin\Theta_p\, \bar H_{p,19}^{(08)}\,,\label{eq: metap}\\
M_{\pi,k}^2 &= \bar H_{p,ii}^{(08)}\quad  i=2,3,4 \,,\\
M_{K,k}^2 &= \bar H_{p,ii}^{(08)}\quad i = 5,6,7,8\,.
\end{align}
In the case of light isospin symmetry we employ
$\bar H_{s/p,19}^{(08)}= \bar H_{s/p,91}^{(08)}$. The non-vanishing
entires of the scalar $\bar H_{s}^{(08)}$ are
{\allowdisplaybreaks
\begin{flalign}
\bar H_{s,11}^{(08)} &= 
-\frac{\bar c}{3} \left(2 \bar\sigma_{L,k}+\sqrt{2}
  \bar\sigma_{S,k}\right)+\frac{1}{3} \bar U^{(2,0)} \left(\sqrt{2}
  \bar \sigma_{L,k}+\bar \sigma_{S,k}\right){}^2+\frac{2}{9} \bar
U^{(0,1)} \left(\bar \sigma_{L,k}-\sqrt{2} \bar
  \sigma_{S,k}\right){}^2 \\ \nonumber 
&\quad+\frac{1}{54}\bar U^{(0,2)} \left(\bar \sigma_{L,k}^2-2 \bar
  \sigma_{S,k}^2\right){}^2 \left(-2 \sqrt{2} \bar \sigma_{L,k} \bar
  \sigma_{S,k}+\bar \sigma_{L,k}^2+2 \bar \sigma_{S,k}^2\right) \\
\nonumber 
&\quad+\frac{1}{9} \bar U^{(1,1)} \left(\bar \sigma_{L,k}^2-2 \bar
  \sigma_{S,k}^2\right) \left(-\sqrt{2}\bar \sigma_{L,k} \bar
  \sigma_{S,k}+2 \bar \sigma_{L,k}^2-2 \bar \sigma_{S,k}^2\right)+
\bar U^{(1,0)} \,, \\ 
\bar H_{s,19}^{(08)} &= \frac{1}{108} \left(\bar \sigma_{L,k}-\sqrt{2} \bar \sigma_{S,k}\right) \bigg[6 \bar U^{(1,1)} \left(5 \bar \sigma_{L,k}^2 \bar \sigma_{S,k}+5 \sqrt{2} \bar \sigma_{L,k} \bar \sigma_{S,k}^2+2 \sqrt{2} \bar \sigma_{L,k}^3+8 \bar \sigma_{S,k}^3\right)+36 \bar U^{(2,0)}
   \left(\sqrt{2} \bar \sigma_{L,k}+\bar \sigma_{S,k}\right) \\ \nonumber
&\quad+18 \sqrt{2} \bar c+\bar U^{(0,2)} \left(\sqrt{2} \bar \sigma_{L,k}+4
   \bar \sigma_{S,k}\right) \left(\bar \sigma_{L,k}^2-2 \bar \sigma_{S,k}^2\right){}^2+6 U^{(0,1)} \left(5 \sqrt{2} \bar \sigma_{L,k}+14
   \bar \sigma_{S,k}\right)\Bigg] \,, \\
\bar H_{s,22}^{(08)} &= \bar c\frac{ \bar\sigma_{S,k}}{\sqrt{2}}+\frac{1}{6} \bar U^{(0,1)} \left(7 \bar\sigma_{L,k}^2-2 \bar\sigma_{S,k}^2\right)+\bar U^{(1,0)} \,, \\
\bar H_{s,55}^{(08)} &=  \frac{\bar c}{2} \bar\sigma_{L,k}+\frac{1}{6}\bar U^{(0,1)} \left(3 \sqrt{2} \bar \sigma_{L,k} \bar \sigma_{S,k}+\bar\sigma_{L,k}^2+4 \bar\sigma
   _{S,k}^2\right)+ \bar U^{(1,0)} \,, \\
\bar H_{s,99}^{(08)} &= \frac{\bar c}{6} \left(4 \bar \sigma_{L,k}-\sqrt{2} \bar \sigma_{S,k}\right)+\frac{1}{18} \bar U^{(0,1)} \left(8 \sqrt{2} \bar \sigma
   _{L,k} \bar \sigma_{S,k}-\bar \sigma_{L,k}^2+22 \bar \sigma_{S,k}^2\right) \\ \nonumber
&\quad +\frac{1}{9} \bar U^{(1,1)} \left(\sqrt{2} \bar \sigma_{L,k} \bar \sigma
   _{S,k}+\bar \sigma_{L,k}^2-4 \bar \sigma_{S,k}^2\right) \left(\bar \sigma_{L,k}^2-2 \bar \sigma_{S,k}^2\right)+\frac{1}{3} \bar U^{(2,0)} \left(-2
   \sqrt{2} \bar \sigma_{L,k} \bar \sigma_{S,k}+\bar \sigma_{L,k}^2+2 \bar \sigma_{S,k}^2\right) \\ \nonumber
&\quad +\frac{1}{108}\bar U^{(0,2)} \left(4 \sqrt{2} \bar \sigma_{L,k} \bar \sigma_{S,k}+\bar \sigma_{L,k}^2+8 \bar \sigma_{S,k}^2\right) \left(\bar \sigma_{L,k}^2-2 \bar \sigma_{S,k}^2\right){}^2 +\bar U^{(1,0)} \,,
\end{flalign}
}
and similarly the non-vanishing entries of the pseudoscalar $H_{p}^{(08)}$ read
{\allowdisplaybreaks
\begin{flalign}
\bar H_{p,11}^{(08)} &= \frac{\bar c}{3} \left(2 \bar \sigma_{L,k}+\sqrt{2} \bar \sigma_{S,k}\right)+\bar U^{(1,0)} \,, \\
\bar H_{p,19}^{(08)} &=  \frac{\bar c}{6} \left(2 \bar \sigma_{S,k}-\sqrt{2} \bar \sigma_{L,k}\right)+\frac{\sqrt{2}}{6} \bar U^{(0,1)} \left(\bar \sigma_{L,k}^2-2
   \bar \sigma_{S,k}^2\right) \,, \label{eq: hp19}\\
\bar H_{p,22}^{(08)} &=  -\frac{\bar c}{\sqrt{2}} \bar \sigma_{S,k}+\frac{1}{6}\bar U^{(0,1)} \left(\bar \sigma_{L,k}^2-2 \bar \sigma_{S,k}^2\right)+
  \bar  U^{(1,0)} \,, \\
\bar H_{p,55}^{(08)} &=  -\frac{\bar c}{2} \bar \sigma_{L,k}+\frac{1}{6}\bar U^{(0,1)} \left(-3 \sqrt{2} \bar \sigma_{L,k} \bar \sigma_{S,k}+\bar \sigma_{L,k}^2+4 \bar \sigma
   _{S,k}^2\right)+ \bar U^{(1,0)} \,, \\
\bar H_{p,99}^{(08)} &=  -\frac{\bar c}{6}\left(4\bar \sigma_{L,k}-\sqrt{2}  \bar\sigma_{S,k}\right)-\frac{1}{6}\bar U^{(0,1)} \left(\bar \sigma_{L,k}^2-2 \bar \sigma
   _{S,k}^2\right)+\bar U^{(1,0)} \,.
\end{flalign}
} The mesonic three-point functions
$\bar \lambda_{\phi_i\phi_j\phi_l,k}$ which enter the meson anomalous
dimension \Eq{eq:etaphi} are defined as
\begin{align}
  \bar \lambda_{\phi_i\phi_j\phi_l,k} = \frac{\partial^3 \bar{\tilde
  U}_k(\bar\Sigma,\bar\Sigma^\dagger)}{\partial \bar\phi_i \partial
  \bar\phi_j \partial \bar\phi_l}\bigg|_{\Phi_0}\,. 
\end{align}
In the following we choose the physical basis where the meson fields
are given by \Eq{eq:mesons}.  Since we define $\eta_{\phi,k}$ from the
wave function renormalization of $\pi^+$, we need the three-point
functions that involve at least one $\pi^+$. We find:
{\allowdisplaybreaks
\begin{align}
\bar\lambda_{\pi^+\pi^-f_0,k} &= \frac{1}{36 \sqrt{3}}
\bigg\{6 \bar U^{(1,1)} \left(\bar \sigma_{L,k}^2-2 \bar \sigma_{S,k}^2\right) \Big[2
   \bar \sigma_{L,k} \left(\cos \Theta _s-\sqrt{2} \sin \Theta _s\right)+\bar \sigma_{S,k} \left(\sin\Theta_s+\sqrt{2} \cos \Theta_s\right)\Big]\\ \nonumber
&\quad-36 \bar U^{(2,0)} \Big[\bar \sigma_{L,k} \left(\sqrt{2} \sin
   \Theta _s-\cos \Theta _s\right)+\bar \sigma_{S,k} \left(\sin \Theta _s+\sqrt{2} \cos\Theta_s\right)\Big]+18\bar c \left(\sqrt{2} \sin \Theta _s+2 \cos \Theta _s\right)\\ \nonumber
&\quad+\bar U^{(0,2)} \left(\bar \sigma_{L,k}^2-2 \bar \sigma_{S,k}^2\right){}^2 \Big[\bar \sigma_{L,k} \left(\cos \Theta _s-\sqrt{2} \sin\Theta_s\right)+2 \bar \sigma_{S,k} \left(\sin \Theta _s+\sqrt{2} \cos \Theta _s\right)\Big]\\ \nonumber
&\quad+12 \bar U^{(0,1)} \Big[\bar \sigma_{L,k} \left(\cos \Theta _s-\sqrt{2} \sin \Theta _s\right)+2 \bar \sigma_{S,k} \left(\sin
   \Theta _s+\sqrt{2} \cos \Theta _s\right)\Big]\bigg\}\,,\\
%%%%%%%%%%%%%%%%
\bar\lambda_{\pi^+\pi^-\sigma,k} &= \frac{1}{36 \sqrt{3}}\bigg\{ 6\bar U^{(1,1)} \left(\bar \sigma_{L,k}^2-2 \bar \sigma_{S,k}^2\right)
   \Big[2 \bar \sigma_{L,k} \left(\sin \Theta _s+\sqrt{2} \cos \Theta _s\right)+\bar \sigma_{S,k} \left(\sqrt{2} \sin
   \Theta _s-\cos \Theta _s\right)\Big]\\ \nonumber
&\quad+36 \bar U^{(2,0)} \Big[\bar \sigma_{L,k} \left(\sin
   \Theta _s+\sqrt{2} \cos \Theta _s\right)+\bar \sigma_{S,k}
  \left(\cos \Theta _s-\sqrt{2} \sin \Theta_s\right)\Big]-18 \bar c
  \left(\sqrt{2} \cos \Theta _s-2 \sin \Theta _s\right)\\ \nonumber 
&\quad+\bar U^{(0,2)} \left(\bar \sigma_{L,k}^2-2\bar 
  \sigma_{S,k}^2\right){}^2 \Big[\bar \sigma_{L,k} \left(\sin \Theta
  _s+\sqrt{2} \cos\Theta_s\right)+2 \bar \sigma_{S,k} \left(\sqrt{2}
  \sin \Theta _s-\cos \Theta _s\right)\Big]\\ \nonumber 
&\quad+12 \bar U^{(0,1)} \Big[\bar \sigma_{L,k} \left(\sin \Theta
  _s+\sqrt{2} \cos \Theta _s\right)+2 \bar \sigma_{S,k} 
   \left(\sqrt{2} \sin \Theta _s-\cos \Theta
  _s\right)\Big]\bigg\}\,,\\ 
%%%%%%%%%%%%%%%%
\bar\lambda_{\pi^+a_0^-\eta,k} &= \frac{1}{\sqrt{6}}\sin\Theta _p
                                 \left(\bar c-2 \bar U^{(0,1)} \bar
                                 \sigma_{L,k}\right)+\frac{1}{\sqrt{3}}
                                 \cos \Theta _p \left(\bar U^{(0,1)}
                                 \bar \sigma_{L,k}+\bar
                                 c\right)\,,\\ 
%%%%%%%%%%%%%%%%
\bar\lambda_{\pi^+a_0^-\eta^\prime,k} &= \frac{1}{\sqrt{3}}\sin\Theta
                                        _p \left(\bar U^{(0,1)} \bar
                                        \sigma_{L,k}+\bar c\right)
                                        -\frac{1}{\sqrt{6}}\cos\Theta
                                        _p \left(\bar c-2 \bar
                                        U^{(0,1)} \bar
                                        \sigma_{L,k}\right)\,,\\ 
%%%%%%%%%%%%%%%%
\bar\lambda_{\pi^+\kappa^-\bar K^0,k} &= \bar\lambda_{\pi^+K^-\bar
                                        \kappa^0,k} = \bar U^{(0,1)}
                                        \bar \sigma_{S,k}+\frac{\bar
                                        c}{\sqrt{2}}\,. 
\end{align}
}

\end{widetext}

\end{appendix}

%%%%%%%%%%%%%%%%%%%%%%%%%%%%%%%%%%%%%%%%%%%%%%%%%%%%%%%%%%%%%%
%%%%%%%%%%%%%%%%%%%%%%%%%%%%%%%%%%%%%%%%%%%%%%%%%%%%%%%%%%%%%%

\bibliography{qcd-phase2}

\end{document}